\documentclass[superscriptaddress,twocolumn,showpacs,pre,amsmath,amssymb]{revtex4}

\usepackage{graphicx}% Include figure files
\usepackage{dcolumn}% Align table columns on decimal point
\usepackage{bm}% bold math
\usepackage{colordvi}
\usepackage{epsfig}
\usepackage{graphicx}
\usepackage{color}

\newcommand{\kap}{\boldsymbol{\kappa}}
\newcommand{\rb}{{\bf r}}
\newcommand{\bp}{\boldsymbol{\partial}}
\newcommand{\qb}{{\bf q}}
\newcommand{\kb}{{\bf k}}
\newcommand{\pb}{{\bf p}}

\newcommand{\sig}{\boldsymbol{\sigma}}

\newcommand{\B}{\boldsymbol{B}}

\newcommand{\R}{\boldsymbol{R}}
\newcommand{\E}{\boldsymbol{E}}

\begin{document}

\title{First-Principles Constitutive Equation for Suspension Rheology}

\author{J.~M.~Brader} 
\affiliation{
Fachbereich Physik, Universit\"at Konstanz, D-78457 Konstanz, Germany}
\altaffiliation{
Present address: Department of Physics, University of Fribourg, CH-1700 Fribourg, Switzerland
}
\author{M.~E.~Cates} 
\affiliation{
SUPA, School of Physics, The University of Edinburgh, 
Mayfield Road, Edinburgh EH9 3JZ, UK
}
\author{M.~Fuchs}
\affiliation{
Fachbereich Physik, Universit\"at Konstanz, D-78457 Konstanz, Germany}


\begin{abstract}
We provide a detailed derivation of a recently developed first-principles 
approach to calculating averages in systems of interacting, spherical 
Brownian particles under time-dependent flow. 
Although we restrict ourselves to flows which are both homogeneous and incompressible, 
the time-dependence and geometry (e.g. shear, extension) are arbitrary. 
The approximations formulated within mode-coupling theory are particularly suited to dense colloidal suspensions and 
capture the slow relaxation arising from particle interactions and the resulting glass 
transition to an amorphous solid. 
The delicate interplay between slow structural relaxation and time-dependent external 
flow in colloidal suspensions may thus be studied within a fully tensorial theory. 
\end{abstract}

\maketitle

\section{Introduction}
Imposing flow on a colloidal suspension distorts the microstructure away 
from that of the quiescent state and induces a nontrivial macroscopic stress response.
The microscopic dynamics underlying this macroscopic behaviour are governed by a 
combination of potential, Brownian and 
hydrodynamic forces which interact in a complicated fashion with the solvent flow field, 
as described by the Smoluchowski equation \cite{dhont,bergenholtz_naegele}.  
The subtle balance between these various physical mechanisms gives rise to a rich 
phenomenology, but also serves to complicate the formulation of tractable theoretical approaches 
\cite{brader_review}. 
In the present work we will present details of a first-principles theory, first outlined in \cite{brader2}, which provides a unified description of the mechanical response of colloidal 
liquids and glasses to external flow, albeit in the absence of hydrodynamic interactions.

Colloidal suspensions are of great importance for practical applications
and the performance of many commercial products and industrial processes often depends 
upon the rheological nonlinearities which occur when colloidal particles are added to a 
Newtonian solvent \cite{coussot}.  
For example, the phenomena of  shear-thinning, shear-thickening and yielding are  
relevant for the even spreading of paints, shock absorption in automobiles and the flow of toothpaste, respectively.
In order to control and tune the rheology of a suspension to meet the needs of a specific 
application it is thus necessary to have an understanding of how the microscopic 
interactions between the constituents influence the macroscopic response \cite{larson1}. 
The challenge to statistical physics is to identify tractable approximation schemes which 
capture the relevant physics while remaining simple enough for concrete calculations to be performed.  

Quiescent monodisperse systems of spherical particles display an 
equilibrium phase diagram with colloidal gas, fluid and crystalline phases, analogous to 
those found in simple fluids \cite{hansen}. 
By introducing a sufficent degree of polydispersity the mechanisms leading to crystallization 
can be suppressed and the system may undergo dynamical arrest to a metastable glass or
gel phase. 
Experimentally, such arrested states are characterized by very slow structural relaxation 
\cite{vanmegan1,vanmegan2,beck}. 
As the system remains nonergodic on measureable timescales it appears as a soft amorphous 
solid with a corresponding elastic modulus. 
Many features of the dynamical arrest observed in colloidal experiments are captured  
by the mode-coupling theory (MCT), both for repulsive glasses for which the arrest is caused 
by steric hinderance \cite{goetze_review} and attractive glasses and gels where interparticle 
attraction dominates the dynamics \cite{pham,zaccarelli} (although ageing dynamics are neglected 
within standard MCT). 

It is fair to say that the MCT-based
theoretical understanding of quiescent glasses is now in a fairly 
advanced state and that the linear response regime is under control. 
However, much less is known about the nonlinear response of arrested states and, despite progress, 
consensus remains to be achieved regarding the fundamental 
physical mechanisms at work.  
A key feature of systems under flow is the competition between 
the timescale of structural relaxation, which becomes very large close to the glass transition, 
and the timescale defined by the inverse of the flow rate. 
This competition is captured in an elegant way by generalized MCT treatments 
\cite{brader1,catesfuchs,fuchs2005,fc09} for which the memory function generating the slow 
relaxation becomes reduced by shear via the mechanism of wavevector advection. 
Various nonlinear rheological phenomena related to shear thinning are found to arise.
The earliest generalizations of MCT to treat nonlinear rheology focused on the special case 
of shear flow \cite{catesfuchs,fuchs2005} (see also \cite{fc09} for a detailed account). 
More recently, the theory has been extended, first to treat time-dependent shear \cite{brader1,zausch} 
and subsequently to treat arbitrary time-dependent flow \cite{brader2}. 
These final developments have elevated the theory to the status of a full constitutive model, 
in the sense that nonsteady three-dimensional flows can be addressed. 
Although full numerical solution of the equations in three-dimensions has not yet been achieved, 
a simplified schematic version of the tensorial theory has been developed which provides sensible 
predictions regarding the flow and yielding behaviour of systems with densities close to and above 
the glass transition \cite{pnas}. 

We would like to clarify the relationship of the present work with our previous publications 
on this topic. 
The original outline for extending MCT to treat nonlinear rheology under steady shear flow was 
presented in the short paper \cite{catesfuchs}. 
Due to the technical difficulty of numerically solving the closed MCT equations in three dimensions, tractable 
schematic and semi-schematic models aiming to capture the essential physical mechanisms were developed 
in \cite{faraday2003}. 
The approach of \cite{catesfuchs} was generalized to treat time-dependent shear flow and 
applied to the special case of shear step-strain \cite{brader1}.
During the subsequent development of the theory we made significant technical improvements which involved 
an alternative definition of the transient density correlator from that employed in 
\cite{catesfuchs,brader1,faraday2003}. (The improved definition is given by (\ref{correlator_def}) below.) 
This new definition ensures that the initial decay rate (see (\ref{initial_decay}) below) remains 
positive for all shear rates, which improves the numerical stability of the theory and leads to 
simpler and more elegant expressions. 
It is this improved formulation which was employed in the short paper \cite{brader2} and which will be 
explained in detail in this work. The simple-shear theory recently presented in \cite{fc09} also employed 
the improved formalism and is a special case of the results presented here. 
It has already been solved numerically for steady shear in two dimensions leading to results for stresses, structure and single particle motion in qualitative agreement with Brownian dynamics simulations \cite{henrich_roy_soc,krueger_epje}.

In this paper we present full details of the tensorial time-dependent theory that was outlined in \cite{brader2}. 
Our theory starts from a well defined microscopic starting point (the Smoluchowski equation) 
and leads directly to an approximation for the time-dependent nonequilibrium distribution
function from which average quantities can be calculated. 
This includes, but is not limited to, an approximation for the macroscopic stress tensor as a 
functional of the velocity gradient tensor.

The paper is structured as follows: 
In Section \ref{microscopic} we define the microscopic dynamics of the colloidal particles. 
In Section \ref{invariance_sec} considerations of translational invariance are employed to 
identify the deformation measures appropriate for describing the system under flow.  
In Section \ref{integrationtt} we develop the integration-through-transients formalism which 
generates a generalized Green-Kubo relation for the macroscopic stress tensor, the mode-coupling 
approximation of which is outlined in Section \ref{constitutive}. 
Our approximate expression for the stress requires knowledge of the transient density correlator. 
In Section \ref{equation_of_motion} we develop a mode-coupling equation of motion for 
this quantity which captures the competition between slow structural relaxation and the fluidizing 
effect of flow. 
In Section \ref{distorted_sec} we consider application of the integration-through-transients 
formalism to calculation of the distorted structure factor and show how an integration over the 
anisotropy, which encodes the microstructural distortion, recovers 
the constitutive equation as obtained earlier in Section \ref{constitutive}.
Finally, in section \ref{discussion} we give a discussion and outlook to future work.

\section{Microscopic dynamics}\label{microscopic}
We consider a system of $N$ spherical Brownian particles of diameter $d$ dispersed in 
a solvent with a specified time dependent velocity profile ${\bf v}(\rb,t)=\kap(t)\cdot\rb$, 
where the velocity gradient $\kap(t)$ is taken to be a fixed input quantity. 
We assume incompressible flow, ${\rm Tr}\,\kap(t)=0$, and will from the outset neglect hydrodynamic 
interactions. 
Our choice to neglect hydrodynamics is largely motivated by the desire to obtain a 
computationally tractable theory, but may also be physically jusifiable for high volume 
fraction states close to the glass transition, where particle motion is slow. 
Given these assumptions the distribution function of particle positions obeys the simplified  
Smoluchowski equation \cite{dhont}
\begin{eqnarray}
\frac{\partial \Psi(t)}{\partial t} &=& \Omega(t) \Psi(t),
\label{smol}
\end{eqnarray}
where the Smoluchowski operator is given by
\begin{eqnarray}
\Omega(t) &=& \sum_{i} \bp_i\cdot[\bp_i - {\bf F}_i - \kap(t)\cdot\rb_i].
\label{smol_op}
\end{eqnarray}
In order to simplify notation we have set both the thermal energy $k_BT$ and bare diffusion coefficient 
$D_0$ equal to unity. 
The total force acting on particle $i$ due to potential interations is given by 
${\bf F}_i=-\bp_i U_N$, 
where $U_N$ is the total potential energy.

\section{Translational invariance}\label{invariance_sec}
Employing a spatially constant velocity gradient has the consequence that the two-time 
correlation functions are invariant with respect to spatial translation. 
%assuming, of 
%course, that there exist no symmetry-breaking nonequilibrium transitions.  
The requirement that translational invariance holds for the two-time correlations 
provides a method to identify the relevant affine deformation measures required for 
our subsequent analysis. 
A prerequisite is an understanding of the invariance properties of the time-dependent 
distribution function $\Psi(t)$.
If the system is taken to be in thermodynamic equilibrium at some initial time $t_0$ which, without 
loss of generality, can be taken as the time origin $t_0=0$, 
the Smoluchowski equation (\ref{smol}) may be formally solved to obtain the 
distribution at later times
\begin{eqnarray}
\Psi(t)=e_+^{\int_{0}^{t}ds\,\Omega(s)}\Psi_e,
\label{ti1}
\end{eqnarray}
where we have introduced a time-ordered exponential function (Appendix \ref{time_ordered}).
In the following we will first use the formal solution (\ref{ti1}) to prove the translational invariance of 
$\Psi(t)$ and then use this result to analyze the two-time correlation functions.

\subsection{Distribution function}\label{sec:distribution}
The fact that $\kap(t)$ does not depend upon spatial coordinates suggests that the translational 
invarance of the equilibrium state will be preserved by the Smoluchowski dynamics; only relative 
particle motion is physically relevant. 
However, proof of this expectation for a general time-dependent flow is complicated by the fact that 
the Smoluchowski operator is itself not translationally invariant.
Shifting all particle coordinates by a constant vector $\rb_i'=\rb_i+{\bf a}$ leads to 
the shifted Smoluchowski operator 
\begin{eqnarray}
\Omega(\Gamma',t)&=&\Omega(\Gamma,t) - {\bf a}\cdot\kap^T(t)\cdot\bf{P} \notag\\
&\equiv&\Omega(\Gamma,t) + A(t),
\label{shift}
\end{eqnarray}
where ${\bf P}\equiv\sum_i \bp_i$ and we have used 
\begin{eqnarray}
{\bf P}\cdot\kap(t)\cdot{\bf a} ={\bf a}\cdot\kap^T(t)\cdot\bf{P}.
\end{eqnarray} 
The positions of all particles are represented by $\Gamma\equiv\{\rb_1,\cdots,\rb_N\}$. 
Substitution of  (\ref{shift}) into (\ref{ti1}) gives the distribution function following the shift
\begin{eqnarray}
\Psi(\Gamma',t)=e_+^{\int_0^t ds\, (\Omega(\Gamma,s)+A(s))}\Psi_e(\Gamma),
\label{shifted}
\end{eqnarray}
where we have used the obvious translational invariance of the equilibrium distribution 
$\Psi_e(\Gamma')=\Psi_e(\Gamma)$. 
The next step is to write the time-ordered exponential explicitly in terms of its series 
definition 
\begin{eqnarray}\label{expansion}
e_+^{\int_0^t ds\, (\Omega(s)+A(s))}\!\!\!&=&\!
1 + \int_{t_1}^{t_2}\!\!\!ds_1 (\Omega(s_1)+A(s_1)) \notag\\
&&\hspace*{-2.5cm}
+ \int_{t_1}^{t_2}\!\!\!ds_1\!\int_{t_1}^{s_1}\!\!\!ds_2\, \big(\Omega(s_1)+A(s_1)\big)
\big(\Omega(s_2)+A(s_2)\big)\notag\\
\notag\\
&&\hspace*{-2.5cm}+\cdots.
\end{eqnarray}
In order to simplify this expression we define the commutators 
\begin{eqnarray}\label{commutators}
c^{(2)}(s_1,s_2) &\equiv& [ A(s_1) , \Omega(s_2) ]
\\
c^{(3)}(s_1,s_2,s_3) &\equiv& [ c^{(2)}(s_1,s_2) , \Omega(s_3) ]
\notag\\
&&\hspace*{-0.4cm}\vdots
\notag\\
c^{(n)}(s_1,\cdots,s_n) &\equiv& [ c^{(n-1)}(s_1,\cdots,s_{n-1}) , \Omega(s_n) ].
\notag
\end{eqnarray}
Due to the fact $A(t)$ is independent of spatial coordinates, simple expressions are 
obtained for the commutators 
\begin{eqnarray}\label{commutators1}
c^{(2)}(s_1,s_2)&=&{\bf a}\cdot\kap^T(s_1)\kap^T(s_2)\cdot \bf{P} 
\notag\\
c^{(3)}(s_1,s_2,s_3)&=&-{\bf a}\cdot\kap^T(s_1)\kap^T(s_2)\kap^T(s_3)\cdot \bf{P} 
\notag\\
&&\hspace*{-0.4cm}\vdots
\notag\\
c^{(n)}(s_1,\cdots,s_n)&=&(-1)^n {\bf a}\cdot\prod_{i=1}^{n}\kap^T(s_{i})\cdot {\bf P}.
\end{eqnarray}
Newton's third law states that the sum of internal forces in the system must be zero.  
It follows that 
\begin{eqnarray}
{\bf P}\psi_e=\sum_{i}\bp_i \frac{e^{-U_N}}{Z}
=\psi_e\sum_{i}{\bf F}_i
=0, 
\end{eqnarray}
where $Z$ is the canonical partition function. 
We thus have
\begin{eqnarray}
A(s_1)\Psi_e(\Gamma)=0 \label{com1}\\
c^{(n)}(s_1,\cdots,s_n)\Psi_e(\Gamma)=0.\label{com2}
\end{eqnarray}
For each term in (\ref{expansion}) the commutation relations (\ref{commutators}) may be used 
to bring all factors of $\Omega$ to the left of the integrand. 
Application of the relations (\ref{com1}) and (\ref{com2}) then eliminates all terms involving 
$\kap$. 
We are thus left with 
\begin{eqnarray}
\hspace*{0.cm}\Psi(\Gamma',t)&=&\!
\Big( 1 + \int_{t_1}^{t_2}\!\!\!ds_1 \Omega(\Gamma,s_1) 
\\
&&\hspace*{0.cm}
+ \int_{t_1}^{t_2}\!\!\!ds_1\!\int_{t_1}^{s_1}\!\!\!ds_2\, \Omega(\Gamma,s_1)
\Omega(\Gamma,s_2)
+\cdots \Big) \Psi_e
\notag\\
&=& \exp_+\!\left(
\int_0^t \!\!\!ds\, \Omega(\Gamma\!,s)
\right)\Psi_e.\notag\\
&=&\Psi(\Gamma,t),
\label{proof}
\end{eqnarray}
which is the desired result. 
We have thus demonstrated that the distibution function is translationally invariant for 
any homogeneous flow field $\kap(t)$ (although it will be anisotropic, reflecting the symmetry 
of the imposed velocity gradient). 

\subsection{Two-time correlation function}
We now consider the properties under uniform translation of the two-time correlation functions; 
namely the correlation of two wavevector dependent fluctuations. 
Without loss of generality we may express the wavevector dependent fluctuations as 
\begin{eqnarray}\label{functions}
f_{\qb}(\Gamma,t,t')=e_-^{\int_{t'}^t ds\,\Omega^{\dagger}(\Gamma,s)}\sum_i X_i^{f}(\Gamma)
e^{i\qb\cdot\rb_i}, 
\end{eqnarray} 
where the adjoint Smoluchowski operator can be obtained from (\ref{smol}) by partial integration
\begin{eqnarray}
\Omega^{\dagger}(t) = \sum_{i} [\,\bp_i + {\bf F}_i + \rb_i\cdot\kap^T(t)\,]\cdot\bp_i\;.
\label{adjoint}
\end{eqnarray}
In contrast to the distribution function, which evolves from time $t'$ to time $t$ according to the propagator 
$\exp_+\!\int_{t'}^t ds\,\Omega(s)$, the time evolution of fluctuations (`observables') is dictated by the adjoint propagator $\exp_-\!\int_{t'}^t ds\,\Omega^{\dagger}(s)$, where the operation of taking the adjoint 
reverses the time ordering (see Eqs.(\ref{adjoint1}) and (\ref{adjoint2})). 
In the present work we will consider only fluctuations for which $X_i^{f}(\Gamma)$ is translationally 
invariant. 
This assumption holds, for example, in the case of the density $\rho_{\qb}(t)$, obtained by setting $X_i^{\rho}=1$ in (\ref{functions}) and the less familiar case of 
wavevector dependent stress fluctuations $\sigma_{\alpha\beta}(\qb,\{ \rb_i \})$ \cite{zoppi,irving}, 
obtained by setting 
\begin{eqnarray}\label{kirkwood}
(X_i^{\sigma})_{\alpha\beta}=\delta_{\alpha\beta} - \frac{1}{2}\sum_{j\ne i} 
\frac{r_{ij}^{\alpha} r_{ij}^{\beta}}{r_{ij}}
\frac{du(r_{ij})}{dr_{ij}}
\end{eqnarray}
in (\ref{functions}). 
According to this definition $\sigma_{\alpha\beta}(\qb,\{ \rb_i \})$ has dimensions of energy (recall that 
we set $k_BT=1$) and thus requires division by a volume to become a true stress. 
One should thus bear in mind that the familiar macroscopic stress arises from taking the average of 
$\sigma_{\alpha\beta}(\qb=0,\{ \rb_i \})/V$. 
 
The two-time correlation function is formally defined by
\begin{eqnarray}
C_{f_{\qb}g^{}_{\kb}}(t,t')
%&=&\langle f^*_{\qb}(t)g^{}_{\kb}(t')\rangle_{\kap(t')}
%\notag
%\\
\label{correlation_formal}
=\int \!\!d\Gamma\,  \Psi(t') 
 f^*_{\bf q}\, e_-^{\int_{t'}^{t}\!ds\,\Omega^{\dagger}(s)} g^{}_{\bf k}.
\end{eqnarray} 
As in the previous subsection, we now shift all particle coordinates by a constant vector, 
$\rb_i\rightarrow\rb_i+{\bf a}$ to obtain the shifted adjoint Smoluchowski operator 
\begin{eqnarray}
\Omega^{\dagger}(\Gamma',t)&=&\Omega^{\dagger}(\Gamma,t) + {\bf a}\cdot\kap^T(t)\cdot {\bf P}\notag\\
&\equiv&\Omega^{\dagger}(\Gamma,t) + A^{\dagger}(t).
\end{eqnarray}
The distribution function appearing in (\ref{correlation_formal}) is invariant with respect to this 
shift as proven in (\ref{proof}).  
To make further progress we employ the identity (\ref{identity2}) to factorize the ordered exponential
\begin{eqnarray}\label{factorize}
e_-^{\int_{t'}^{t}\!ds\,\Omega^{\dagger}(\Gamma,s) + A^{\dagger}(s)}&=&
e_-^{\int_{t'}^{t}\!\!ds\; \Omega^{\dagger}(\Gamma,s)}\times\\
&&\hspace*{-3cm}\times\exp_-\left(\int_{t'}^{t}\!\!ds\; e_+^{-\int_s^{t}ds'\Omega^{\dagger}(\Gamma,s')}
A^{\dagger}(s)\,e_-^{\int_s^{t}ds'\Omega^{\dagger}(\Gamma,s')}\right).
\notag
\end{eqnarray}
The argument of the second exponential is a complicated `interaction representation' of the operator 
$A^{\dagger}(s)$, but can be simplified considerably using the nested commutator expansion 
(\ref{hadamard1}) derived in Appendix \ref{hadamard}. 
Defining the commutators 
\begin{eqnarray}\label{newcom1}
b^{(2)}(s_1,s_2)&\equiv& [\Omega^{\dagger}(s_1),A^{\dagger}(s_2)] 
\\
\hspace*{0cm}
b^{(3)}(s_1,s_2,s_3)&\equiv& [\Omega^{\dagger}(s_1),[\Omega^{\dagger}(s_2),A^{\dagger}(s_3)]] 
\notag\\
\hspace*{0cm}
b^{(4)}(s_1,s_2,s_3,s_4)&\equiv& 
[\Omega^{\dagger}(s_1),[\Omega^{\dagger}(s_2),[\Omega^{\dagger}(s_3),A^{\dagger}(s_4)]]] 
\notag\\
&\vdots&
\notag
\end{eqnarray}
we find that 
\begin{eqnarray}\label{newcom2}
b^{(2)}(s_1,s_2)&=&-{\bf a}\cdot\kap^T(s_2)\kap^T(s_1)\cdot \bf{P} 
\notag\\
b^{(3)}(s_1,s_2,s_3)&=&{\bf a}\cdot\kap^T(s_3)\kap^T(s_2)\kap^T(s_1)\cdot \bf{P} 
\notag\\
&&\hspace*{-0.4cm}\vdots
\\
b^{(n)}(s_1,\cdots,s_n)&=&(-1)^{(n+1)} {\bf a}\cdot\prod_{i=0}^{n-1}\kap^T(s_{n-i})\cdot {\bf P}.
\notag
\end{eqnarray}
Using (\ref{newcom1}) and (\ref{newcom2}) in the expansion (\ref{hadamard1}) yields 
a simplification of the interaction representation operator
\begin{eqnarray}
&&\hspace*{-0.5cm} e_+^{-\int_s^{t}ds'\Omega^{\dagger}(\Gamma,s')}
A^{\dagger}(s)\,e_-^{\int_s^{t}ds'\Omega^{\dagger}(\Gamma,s')}
\notag\\
&&\hspace*{0.cm}={\bf a}\cdot\kap^T(s)\Big(
{\bf 1} + \int_s^t \!\!ds_1\,\kap^T(s_1) 
\notag\\
&&\hspace*{0.6cm}+ \int_s^t \!\!ds_1\! \int_s^{s_1} \!\!\!\!ds_2\,\kap^T(s_2)\kap^T(s_1)
+\cdots\Big)\cdot{\bf P}
\notag\\
&&=-{\bf a}\cdot\left(\frac{\partial}{\partial s}\,e_-^{\int_s^{t}ds' \kap^T(s')}
\right)\cdot{\bf P}
\end{eqnarray}
where the last equality follows from (\ref{deriv4}) and the definition of the ordered exponential 
(\ref{eminus}).
Substitution into (\ref{factorize}) yields the useful result
\begin{eqnarray}\label{factorize_simplified}
e_-^{\int_{t'}^{t}\!ds\,\Omega^{\dagger}(\Gamma,s) + A^{\dagger}(s)}&=&
e_-^{\int_{t'}^{t}\!\!ds\; \Omega^{\dagger}(\Gamma,s)}\times\\
&&\hspace*{-3cm}\times\exp\left(
{\bf a}\cdot \left(  -{\bf 1} +  
e_-^{\int_{t'}^{t}\!ds\,\kap^T\!(s)}
\right)\cdot{\bf P}
%\int_{t'}^{t}\!\!ds
\right),
\notag
\end{eqnarray}
Substitution of (\ref{factorize_simplified}) into (\ref{correlation_formal}), noting that 
${\bf P}\,\! g^{}_{\kb}=i\,\kb\,\! g^{}_{\kb}$ and using the adjoint relation 
(\ref{adjoint2}) leads to our final result
\begin{equation}
C_{f_{\qb}g^{}_{\kb}}(t,t') = e^{-i (\, \qb - \kb(t,t')\,)\cdot {\bf a}}
\,C_{f_{\qb}g^{}_{\kb}}(t,t'),
\label{invariance}
\end{equation}
where the time-dependent wavevector is given by
\begin{eqnarray}
\kb(t,t')=\kb\cdot e_+^{\int_{t'}^{t}\!ds\,\kap(s)},
\label{advection2}
\end{eqnarray}
The condition of translational invariance thus requires that the phase factor vanishes and 
leads us the conclusion that fluctuation at wavevector $\kb$ at earlier time $t'$ is correlated 
with one at $\qb=\kb(t,t')$ at later time $t$, as a result of the affine flow. 
Equivalently, one can employ the inverse transformation (see Appendix \ref{time_ordered}) 
to show that an earlier fluctuation at wavevector $\kb=\bar{\qb}(t,t')$ is correlated 
with a later one at wavevector $\qb$, where 
\begin{eqnarray}
\bar{\qb}(t,t')=\qb\cdot e_-^{-\int_{t'}^{t}\!ds\,\kap(s)}.
\label{advection1}
\end{eqnarray} 
In the following we will refer to (\ref{advection1}) as the forward-advected wavevector and 
(\ref{advection2}) as the reverse-advected wavevector.

\subsection{Rheological tensors and advection}\label{continuum}
The ordered exponential identified in (\ref{advection1}) may appear unfamiliar, but is nothing more than 
the inverse of the deformation gradient tensor 
\begin{eqnarray}\label{deformation}
\E(t,t')=\frac{\partial \rb(t)}{\partial \rb(t')},
\end{eqnarray}
a standard quantity in elasticity theory \cite{larson,hassager}. 
The deformation gradient transforms a vector (`material line') at time $t'$ 
to a new vector at later time $t$. 
Thus, for spatially homogeneous deformations $\rb(t)$ is transformed from the past to the present via
\begin{eqnarray}
\rb(t)=\E(t,t')\cdot\rb(t'), 
\end{eqnarray}
where 
$E_{\alpha\beta}=\partial r_{\alpha}/\partial r_{\beta}$, and from the present to the past using 
the inverse deformation gradient
\begin{eqnarray}
\rb(t')=\E^{-1}(t,t')\cdot\rb(t). 
\end{eqnarray} 
Taking the time derivative of (\ref{deformation}) and applying the chain rule to the 
right hand side generates a differential equation connecting the deformation to the velocity gradient 
\begin{eqnarray}
\frac{\partial}{\partial t}\E(t,t')=\kap(t)\E(t,t').
\label{velocity_gradient}
\end{eqnarray}
Given the boundary condition $E(t,t)=1$ 
the general solution of (\ref{velocity_gradient}) is thus
\begin{eqnarray}
\E(t,t')=e_+^{\int_{t'}^{t}ds\,\kap(s)},
\label{def}
\end{eqnarray}
with inverse (see Appendix \ref{time_ordered})
\begin{eqnarray}
\E^{-1}(t,t')=e_-^{-\int_{t'}^{t}ds\,\kap(s)}.
\label{invdef}
\end{eqnarray}
The tensors (\ref{def}) and (\ref{invdef}) are precisely those identified from our consideration 
of the translational invariance of wavevector dependent density fluctuations, namely 
$\bar{\kb}(t,t')=\kb\cdot\E^{-1}(t,t')$ and $\kb(t,t')=\kb\cdot\E(t,t')$. 
%We thus have the interpretation that a wavevector in the past is transformed to the present 
%according to 
%\begin{eqnarray}
%\qb(t)=\qb(t')\cdot \E(t,t'),
%\end{eqnarray}
%whereas the current wavevector is transformed to that at an earlier time by 
%\begin{eqnarray}
%\qb(t')=\qb(t)\cdot \E^{-1}(t,t').
%\end{eqnarray}

\section{Integration through transients}\label{integrationtt}
\subsection{Nonequilibrium distribution function}
If we now assume that the system was in thermodynamic equilibrium in the infinite past 
the formal solution (\ref{ti1}) becomes
\begin{eqnarray}
\Psi(t)=e_+^{\int_{-\infty}^{t}ds\,\Omega(s)}\Psi_e.
\label{eplus_sol}
\end{eqnarray}
Despite being formally exact, the fact that the time-evolution is entirely contained within the 
distribution function proves inconvenient when attempting to make a closure approximation. 
For this reason we seek an alternative, but equivalent, solution for which the observables evolve 
in time and the distribution function remains fixed. 
The procedure is analogous to the passage from Schr\"odinger to Heisenberg pictures 
in quantum mechanics and will lead to a Dyson-equation like representation of the transient dynamics 
\cite{evans_morriss}.

Without loss of generality we can separate both the Smoluchowski operator and the distribution 
function into equilibrium and nonequilibrium contributions
\begin{eqnarray}
\Omega(t) &=& \Omega_e + \delta\Omega(t)
\label{sep1}\\
\Psi(t) &=& \Psi_e + \delta\Psi(t),
\label{sep2}
\end{eqnarray} 
where $\delta\Omega(t)=-\sum_{i}\bp_i\cdot(\kap(t)\cdot\rb_i)$ and $\delta\Psi(t)$ 
remains to be determined. 
The equilibrium distribution satisfies $\Omega_e\Psi_e=0$, whereas nonequilibrium steady states 
are determined by $\Omega\Psi=0$. 
When acting upon the equilibrium distribution the nonequilibrium Smoluchowski operator generates 
a term proportional to the stress tensor
\begin{eqnarray}
\Omega(t)\Psi_e=\delta\Omega(t)\Psi_e= (\kap(t)\!:\!\hat\sig) \Psi_e,
\end{eqnarray}
where ${\bf A}:{\bf B}\equiv \sum_{\alpha\beta}A_{\alpha\beta}B_{\beta\alpha}$ and 
$\hat\sig$ is the zero wavevector limit of the potential part of the stress tensor, which we now 
write in the following form
\begin{eqnarray}
\hat\sigma_{\alpha\beta} = \delta_{\alpha\beta} -\sum_{i}r^{\alpha}_iF_i^{\beta}.
\label{micro_stress}
\end{eqnarray}
In order to avoid any confusion we note that a simple transformation to relative coordinates 
suffices to demonstrate the equivalence of (\ref{micro_stress}) to the more familiar Kirkwood form
(\ref{kirkwood}) \cite{irving}. 
Thus, despite first impressions, the compact expression (\ref{micro_stress}) to be employed in our 
calculations does not depend upon the choice of coordinate origin. 

For the incompressible flow under consideration we have 
$\kap(t)\!:\!\boldsymbol{\delta}={\rm Tr}\,\kap(t)=0$, such that the Kronecker delta in 
(\ref{micro_stress}) is irrelevant at this stage of the calculation.
Substitution of (\ref{sep1}) and (\ref{sep2}) into (\ref{smol}) thus yields an equation for 
$\delta\Psi(t)$ 
\begin{eqnarray}
\frac{\partial }{\partial t}\delta\Psi(t) &=& \Omega(t) \delta\Psi(t) 
+ (\kap(t)\!:\!\hat\sig) \Psi_e,
\label{deltapsi}
\end{eqnarray}
which is a simple first order differential equation with inhomogeneity 
$(\kap(t)\!:\!\hat\sig) \Psi_e$.
Assuming an equilibrium distribution in the infinite past, (\ref{deltapsi}) 
is solved by
\begin{eqnarray}
\delta\Psi(t) = \int_{-\infty}^{t}\!\!\!\!\! dt_1\,
e_+^{\int_{t_1}^{t}ds\,\Omega(s)} 
(\kap(t_1)\!:\!\hat\sig) \Psi_e,
\label{delta_sol}
\end{eqnarray}
as can be verified by either variation of parameters or the method of Greens functions. 
The full nonequilibrium distribution function is given by substitution of 
(\ref{delta_sol}) into (\ref{sep2}). 
The nonequilibrium average of an arbitrary function of the particle coordinates 
$f(\{\rb^N\})$, which need not be a scalar quantity, is thus given by
\begin{eqnarray}
\langle f\rangle^{\rm ne} = \langle f \rangle
+ \int \!\!d\Gamma\!\! \int_{-\infty}^{t}\!\!\!\!\! dt_1\, f\,
e_+^{\int_{t_1}^{t}ds\,\Omega(s)} 
(\kap(t_1)\!:\!\hat\sig) \Psi_e,
\label{average1}
\end{eqnarray}
where the integral over $\Gamma$ is a phase-space integral over all particle coordinates and 
$\langle\cdot\rangle$ denotes an equilibrium average.
The final result, central to the present theoretical development, is obtained by 
partial integration
\begin{eqnarray}
\langle f\rangle^{\rm ne} = \langle f \rangle
+ \int_{-\infty}^{t}\!\!\!\!\! dt_1\,
\langle
\kap(t_1)\!:\!\hat\sig\, 
e_-^{\int_{t_1}^{t}ds\,\Omega^{\dagger}(s)} f\rangle.
\label{average2}
\end{eqnarray}
The time development of the observable $f$ is thus generated by the adjoint operator.

As the test function $f$ is arbitrary, it may be removed to provide a formal operator 
expression for the distribution function  
\begin{eqnarray}
\Psi(t) &=& \Psi_e + \int_{-\infty}^{t}\!\!\!\!\! dt_1\,
\Psi_e\, \kap(t_1)\!:\!\hat\sig \,e_-^{\int_{t_1}^{t}ds\,\Omega^{\dagger}(s)}. 
\label{itt}
\end{eqnarray}
This expression is equivalent to (\ref{eplus_sol}) and implies that the function 
to be averaged should be multiplied by (\ref{itt}) from the left and integrated over the particle
coordinates. 
Equation (\ref{itt}) is the central result of the integration through transients formalism 
and we will demonstrate that this provides a very convenient starting point for analysis of 
the nonequilibrium dynamics of colloidal suspensions.
We note that analogous expressions for the nonequilibrium distribution function have been 
considered by Evans and Morriss \cite{evans_morriss} and Chong and Kim \cite{chong}, based on the 
thermostatted SLLOD equations of motion.  
A fundamental advantage of (\ref{itt}) over (\ref{eplus_sol}) is that it enables nonequilibrium 
averages to be expressed in terms of equilibrium averages, under the sole
assumption that the system was in equilibrium in the infinite past. 
This leads to a Dyson-like representation of the dynamics, reminiscent of time-dependent 
perturbation theory in quantum mechanics.
Within our formalism a range of nonergodic quiescent states can be generated from this initial 
state by imposing various flow histories. 
This includes the typical experimental protocol for which strong pre-shear is applied in order 
to erase memory and obtain a reproducible initial state from which the system age may be measured.

\subsection{Exact projection}
A potential pitfall of applying (\ref{itt}) arises from the existence of conservation laws 
which lead to zero eigenvalues of $\Omega^{\dagger}$ and which could, in principle, 
prevent convergence of the time integral. 
In the following we will construct an exact reformulation of (\ref{average2}) in which slow fluctuations 
are explicitly projected out, ensuring convergence of the integral for all values of $\qb$ and thus 
preparing the ground for subsequent approximations.

For the Brownian dynamics under consideration only the particle number is conserved,  
$\partial_t \rho_{\qb=0}=\Omega^{\dagger}(t) \rho_{\qb=0}=0$, which suggests a possible divergence in the 
hydrodynamic limit when the arbitrary test function in (\ref{average2}) is chosen to be a density 
fluctuation, $f=\rho_{\qb}$. 
Fortunately, this poses no difficulty, as the average in the integrand vanishes for all wavevectors
\begin{eqnarray}
\langle\kap(t)\!:\!\hat\sig \,e_-^{\int_{t_1}^{t}ds\,\Omega^{\dagger}(s)} \rho_{\qb}\rangle=0,
\label{orthogonal}
\end{eqnarray}
which ensures that density fluctuations do not couple linearly to the nonequilibrium part of the 
distribution function. 
The fact that (\ref{orthogonal}) holds at $q=0$ follows directly from inserting $\rho_{\qb=0}=N$ 
\begin{eqnarray}
\langle\kap(t)\!:\!\hat\sig \,e_-^{\int_{t_1}^{t}ds\,\Omega^{\dagger}(s)} N\rangle
=
N\langle\kap(t)\!:\!\hat\sig\rangle=0,
\label{easy}
\end{eqnarray}
where the second equality relies on the condition ${\rm Tr}\,\kap(t)=0$. 
For nonzero $q$ the vanishing of the average in (\ref{orthogonal}) is a consequence of 
translational invariance. 
Shifting all coordinates by a constant vector 
$\rb'_i=\rb_i+{\bf a}$ and applying (\ref{factorize_simplified}) leads to
\begin{eqnarray}
\langle\kap(t)\!:\!\hat\sig \,e_-^{\int_{t_1}^{t}ds\,\Omega^{\dagger}(\Gamma',s)} \rho_{\qb}\rangle
&=&\\
&&\hspace*{-4cm}\exp\left(-i\qb\cdot e_-^{-\int_{t'}^{t}\!ds\,\kap(s)}\right)
\langle\kap(t)\!:\!\hat\sig \,e_-^{\int_{t_1}^{t}ds\,\Omega^{\dagger}(\Gamma,s)} \rho_{\qb}\rangle.
\notag
\end{eqnarray}
To prevent violation of translational invariance the average must be identically zero. 
Eq.(\ref{orthogonal}) is thus proven.

We now introduce the density projector $P$ and complement $Q$ 
\begin{eqnarray}
P=\sum_{\kb} |\rho_{\qb}\rangle\frac{1}{N S_q}\langle \rho^*_{\qb}|\,,
\hspace*{1cm}
Q=1-P\,,
\label{1projector}
\end{eqnarray}
where $S_q$ is the static structure factor $S_q=\langle\rho^{*}_{\qb}\,\rho^{}_{\qb}\rangle/N$. 
In (\ref{1projector}) we have introduced a bra-ket notation for wavector dependent fluctuations. 
The scalar product of two arbitrary variables is defined by an equilibrium average 
\begin{eqnarray}\label{braket}
\langle f^*_{\qb}\,|\,g^{}_{\qb}\rangle
\equiv
\int \!d\Gamma\,
\Psi_e f^*_{\qb}(\Gamma)g_{\qb}(\Gamma).
\end{eqnarray}
The projectors (\ref{1projector}) are both idempotent and orthogonal 
\begin{eqnarray}
P^2=P, \hspace*{0.8cm} Q^2=Q, \hspace*{0.8cm} QP=0, 
\end{eqnarray}
and by construction satisfy 
\begin{eqnarray}
P|\rho_{\qb}\rangle=|\rho_{\qb}\rangle  \hspace*{1cm} Q|\rho_{\qb}\rangle=0.
\end{eqnarray}
A direct consequence of (\ref{orthogonal}) is that 
our central result (\ref{average2}) may be exactly rewritten as
\begin{eqnarray}
\langle f\rangle^{\rm ne} = \langle f \rangle
+ \int_{-\infty}^{t}\!\!\!\!\! dt_1\,
\langle
\kap(t_1)\!:\!\hat\sig\,Q 
e_-^{\int_{t_1}^{t}ds\,\Omega^{\dagger}(s)}\,Q f\rangle,
\label{average_QQ}
\end{eqnarray}
where we have again used $\langle\kap(t)\!:\!\hat\sig\rangle=0$.
To proceed further we consider a 
time-dependent generalization of a standard operator identity 
(see e.g. \cite{zwanzig} for the standard result)
\begin{eqnarray}\label{zwanzig_identity}
e_-^{\int_{t_1}^t \!ds\, \Omega^{\dagger}(s)}
&=& 
e_-^{\int_{t_1}^t\! ds\, Q\,\Omega^{\dagger}(s)}\\
&&\hspace*{-1cm}+ 
\int_{t_1}^{t}\! ds'\, e_-^{\int_{t_1}^{s'}\!ds\, \Omega^{\dagger}(s)}  P\,\Omega^{\dagger}(s')
e_-^{\int_{s'}^{t}\!ds\,  Q\,\Omega^{\dagger}(s)},
\notag
\end{eqnarray}
which can easily be verified by differentiation (applying the rules from Appendix \ref{time_ordered}).
Substitution of this identity into (\ref{average_QQ}) and using (\ref{orthogonal}) eliminates the 
complicated second term in (\ref{zwanzig_identity}). 
Finally, exploiting the idempotency of $Q$ enables us to write 
\begin{eqnarray}
e_-^{\int_{t_1}^t\! ds\, Q\,\Omega^{\dagger}(s)}Q
=
e_-^{\int_{t_1}^t\! ds\, Q\,\Omega^{\dagger}(s)\,Q}Q,
\end{eqnarray}
which leads to our final result
\begin{eqnarray}
\hspace*{-0.1cm}\langle f\rangle^{\rm ne}\!\! =\! \langle f \rangle
+\! \int_{-\infty}^{t}\!\!\!\!\! dt_1\,
\langle
\kap(t_1)\!:\!\hat\sig\,Q 
e_-^{\int_{t_1}^{t}ds\,Q\,\Omega^{\dagger}(s)\,Q}\,Q f\rangle.
\label{average_QQQQ}
\end{eqnarray}
The projected expression (\ref{average_QQQQ}) is formally equivalent to (\ref{average2}). 
The advantage of this reformulation is that the absence 
of linear coupling to density fluctuations is explicit in the projected time evolution operator, 
which ensures that approximations to the equilibrium average in (\ref{average2}) will not introduce 
spurious linear couplings.  
It should be noted that these considerations rely on the incompressibility condition ${\rm Tr}\,\kap=0$, 
the relaxation of which would indeed lead to a finite linear coupling for some compressible flows.

\section{Constitutive equation}\label{constitutive}
\subsection{Generalized Green-Kubo formula}
In order to address the rheology of colloidal suspensions we now consider application of 
(\ref{itt}) to calculation of the macroscopic stress tensor. 
Replacing $f$ in (\ref{average2}) by the fluctuating stress tensor elements (\ref{micro_stress}) 
yields a formally exact constitutive equation
\begin{eqnarray}
\sig(t) = \frac{1}{V}\int_{-\infty}^{t}\!\!\!\!\! dt_1\,
\langle\kap(t_1)\!:\!\hat\sig \,e_-^{\int_{t_1}^{t}ds\,\Omega^{\dagger}(s)} \hat\sig\rangle.
\label{stress_general}
\end{eqnarray}
The stress tensor is thus a nonlinear functional of the velocity gradient tensor as $\kap(t)$ 
appears in the adjoint Smoluchowski operator. 
Equation (\ref{stress_general}) thus goes beyond the standard Green-Kubo formulae of linear response 
theory and makes possible a theoretical study of nonlinear rheology. 
\par
The general result (\ref{stress_general}) shows that for arbitrary time-dependent flow the integrand 
must depend upon two time arguments, rather 
than a simple  time  difference. 
To illustrate this point, consider the special case of simple shear flow 
($\kap_{\alpha\beta}(t)=\dot\gamma(t)\delta_{x\alpha}\delta_{y\beta}$, where $\dot\gamma(t)$ is the 
shear rate). 
The shear stress is given by
\begin{eqnarray}
\sigma_{xy}(t) &=& \int_{-\infty}^{t}\!\!\!\!\! dt_1\,
\dot\gamma(t_1)
\left(\frac{1}{V}
\langle\,\hat{\sigma}_{xy}e_-^{\int_{t_1}^{t}ds\,\Omega^{\dagger}(s)} 
\hat\sigma_{xy}\rangle \right)
\notag\\
&\equiv& \int_{-\infty}^{t}\!\!\!\!\! dt_1\,
\dot\gamma(t_1)\,G(t,t_1;[\kap]),
\label{stress_exact_shear}
\end{eqnarray}
where the second equality serves to define the nonlinear shear modulus, which 
is a functional of the velocity gradient tensor. 
The lack of time translational invariance demonstrated by $G(t,t';[\kap])$ is a general feature of 
two-time correlation functions in systems under time-dependent shear and is unavoidable if one 
wishes to go beyond linear response \cite{brader_osc}. 
Time translational invariance can only be recovered by replacing in (\ref{stress_exact_shear}) 
the full evolution operator $\Omega^{\dagger}(t)$ with the time-independent equilibrium 
operator. 
For notational convenience we will henceforth employ $G(t,t')\equiv G(t,t';[\kap])$, omitting explicit 
reference to the functional dependence of the modulus on the velocity gradient.

\subsection{Mode-coupling approximation}
The exact generalized Green-Kubo expression for the stress tensor (\ref{stress_general}) 
requires approximation before explicit calculations can be performed. 
We begin by exactly re-expressing (\ref{stress_general}) in terms of the projected dynamics 
(\ref{average_QQQQ})
\begin{eqnarray}
\sig(t) = \frac{1}{V}\int_{-\infty}^{t}\!\!\!\!\! dt_1\,
\langle\kap(t_1)\!:\!\hat\sig\, Q \,e_-^{\int_{t_1}^{t}\!ds\,Q\,\Omega^{\dagger}(s)\,Q} Q\,\hat\sig\rangle,  
\label{stress_general_proj}
\end{eqnarray}
which will be approximated by considering the overlap of stress fluctuations 
with the simplest relevant slow fluctuations. 
Due to the projector $Q$ in (\ref{stress_general_proj}) the lowest nonzero order in fluctuation 
amplitude is taken to be $\rho^{}_{\kb}\rho^{}_{\pb}$, where $\kb$ and $\pb$ are two distinct wavevectors. 
We thus define a projection operator onto density pairs 
\begin{eqnarray}\label{pair_projector}
P_2=\sum_{\kb > \pb} |\,\rho^{}_{\kb}\rho^{}_{\pb}\rangle 
\frac{1}{N^2S_k S_p} \langle\rho^{*}_{\kb}\rho^{*}_{\pb}|,
\end{eqnarray}
which we assume will capture the dominant slow fluctuations. 
It should be pointed out that this projection operator is only approximate, as we have 
employed the Gaussian ansatz 
\begin{eqnarray}\label{gaussian}
\langle \rho^{*}_{\kb}\rho^{*}_{\pb}\rho^{}_{\kb}\rho^{}_{\pb} \rangle
\approx
\langle \rho^{*}_{\kb}\rho^{}_{\kb} \rangle \langle \rho^{*}_{\pb}\rho^{}_{\pb} \rangle
=N^2S_k S_p
\end{eqnarray}
in the denominator \cite{goetze}. 
Although the approximate projector (\ref{pair_projector}) is not perfectly idempotent, it has the 
advantage that the normalization is expressed in terms of easily calculable equilibrium structure 
factors.    
%and a pair projector onto reverse-advected fluctuations 
%\begin{eqnarray}
%P'_2=\sum_{\kb > \pb} |\rho^{}_{\kb(t,t')}\rho^{}_{\pb(t,t')}\rangle 
%\frac{1}{N^2S_k S_p} \langle\rho^{*}_{\kb(t,t')}\rho^{*}_{\pb(t,t')}|,
%\end{eqnarray}
%where the reverse-advected wavevector is given by (\ref{advection2}). 
Projecting onto density pairs yields an approximation to the stress 
\begin{eqnarray}
\sig(t) \!\!&=&\!\! \frac{1}{V}\!\int_{-\infty}^{t}\!\!\!\!\!\! dt_1\,
\langle\kap(t_1)\!:\!\hat\sig\, Q P_2\,e_-^{\int_{t_1}^{t}\!ds\,Q\,\Omega^{\dagger}(s)\,Q}
P_2\, Q\,\hat\sig\rangle,
\notag\\
&&\hspace*{-1.1cm}=\!\!\!\!\!\mathop{\sum_{\kb > \pb}}_{\;\;\,\kb' > \pb'}\!\!\!
\frac{1}{VN^4}\!\!\int_{-\infty}^{t}\!\!\!\!\!\! dt_1\, 
\frac{V^{(1)}_{\kb'\pb'}
V^{(2)}_{\kb\pb}}
{S_{k'} S_{p'}S_k S_p}
\langle\rho^{*}_{\kb'}\rho^{*}_{\pb'}
\,e_-^{\int_{t_1}^{t}\!ds\,Q\,\Omega^{\dagger}(s)\,Q}
\!\rho^{}_{\kb}\rho^{}_{\pb}\rangle
\notag\\
\label{stress_general_proj1}
\end{eqnarray}
where the vertex functions are given by 
\begin{eqnarray}
V^{(1)}_{\kb'\pb'}&=&\langle\kap(t_1)\!:\!\hat\sig\, Q \rho^{}_{\kb'}\rho^{}_{\pb'} \rangle
\label{trace_vertex}
\\
V^{(2)}_{\kb\pb}&=&\langle\rho^{*}_{\kb}\rho^{*}_{\pb} Q\, \hat\sig\rangle.
\label{dyadic_vertex}
\end{eqnarray}
The second vertex is a tensor, whereas the first is a scalar because of the contraction 
with the velocity gradient.
A straightforward calculation enables the vertices to be expressed in terms of the 
equilibrium static structure factor 
\begin{eqnarray}\label{vertex1}
V^{(1)}_{\kb'\pb'}&=&
\kap(t_1)\!:\!\kb'\pb'\,\frac{1}{k}\frac{dS_{k'}}{dk'}\,\delta_{\kb',-\pb'}
\\
\label{vertex2}
V^{(2)}_{\kb'\pb'}&=&\kb\pb\frac{1}{k}\frac{dS_{k}}{dk}\,\delta_{\kb,-\pb},
\end{eqnarray}
where $\kb\pb$ indicates a dyadic product. 
The wavevector restrictions imposed by the Kronecker deltas appearing in (\ref{vertex1}) 
and (\ref{vertex2}) reduce the fourfold sum in (\ref{stress_general_proj1}) to a double sum. 
We are thus led to consider the four point correlator 
\begin{eqnarray} 
\langle\rho^{*}_{\kb'}\rho^{*}_{-\kb'}
\,e_-^{\int_{t_1}^{t}\!ds\,Q\,\Omega^{\dagger}(s)\,Q}
\!\rho^{}_{\kb}\rho^{}_{-\kb}\rangle.
\end{eqnarray} 
In the spirit of quiescent mode-coupling theory we approximate the unknown four-point density 
correlator by a product of pair correlators and replace the Q-projected dynamics by the full dynamics
\begin{eqnarray}\label{kawasaki}
\langle\rho^{*}_{\kb'}\rho^{*}_{-\kb'}
\,e_-^{\int_{t_1}^{t}\!ds\,Q\,\Omega^{\dagger}(s)\,Q}
\rho^{}_{\kb}\rho^{}_{-\kb}\rangle
&\approx&
\\
&&\hspace*{-5.25cm}
\langle\rho^{*}_{\kb'}
\,e_-^{\int_{t_1}^{t}\!ds\,\Omega^{\dagger}(s)}
\rho^{}_{\kb}\rangle
\langle\rho^{*}_{\kb'}
\,e_-^{\int_{t_1}^{t}\!ds\,\Omega^{\dagger}(s)}
\rho^{}_{\kb}\rangle.
\notag
\end{eqnarray}
For quiescent states this approximate step is well established and amounts to assuming that the density 
fluctuations are Gaussian random variables \cite{sjoegren}. 
We assume here that this approximation remains valid under flow.

The essential observation at this point of the calculation is that the wavevectors $\kb$ and 
$\kb'$ appearing in (\ref{stress_general_proj1}) are not independent. 
The static structure factor and, consequently, the vertices (\ref{vertex1}) and (\ref{vertex2}) 
are clearly invariant with respect to spatial translation. 
In order for the stress (\ref{stress_general_proj1}) to remain translationally invariant we must 
therefore impose that the mode-coupling factorization of the four-point correlator (\ref{kawasaki}) also 
remain invariant. It is this requirement which couples the wavevectors $\kb$ and $\kb'$.
Shifting all particle coordinates by a constant vector and application of (\ref{factorize_simplified}) 
reveals that the the right hand side of (\ref{kawasaki}) is translationally invariant only if 
$\kb'=\kb(t,t')$, where the reverse-advected wavevector is given by (\ref{advection2}). 
This identification leads to the appearance of $\kb(t,t')$ at several places within 
(\ref{stress_general_proj1}) and thus generates a highly nonlinear functional dependence 
of the stress on the velocity gradient tensor.

Substitution of (\ref{vertex1}), (\ref{vertex2}) and (\ref{kawasaki}) into (\ref{stress_general_proj1}), 
using the identification $\kb'=\kb(t,t')$ and replacing the remaining discrete wavevector sum by an 
integral yields 
\begin{eqnarray}\label{constit1}
      \sig(t)\!&=&\!\!\int_{-\infty}^{t} \!\!\!\!\!\!dt'\int\!\!\! 
      \frac{d{\bf k}}{16\pi^3}\!\left(\frac{\kb(t,t')\!\cdot\!
      \kap(t')
      \!\cdot\kb(t,t')}{k k(t,t')}\!\right)\times 
      \notag\\
      &&\hspace*{1cm}\times\;\kb\kb
      \frac{S'_{k}S'_{k(t,t')}}{S^2_{k}}\,\,
      \Phi^2_{\kb (t,t')}(t,t'),
\end{eqnarray}
where we have defined the transient density correlator
\begin{eqnarray}\label{correlator_def}
\Phi_{\bf k}(t,t') = \frac{1}{NS_{k}}\langle\, \rho^*_{\bf k}\,
e_-^{\int_{t'}^t ds\,\Omega^{\dagger}(s)}
\rho_{\bar{\kb}(t,t')} \rangle,
\end{eqnarray}
which remains to be determined. Note that the forward-advected wavevector appears naturally in 
(\ref{correlator_def}) as a result of 
\begin{eqnarray}
\Phi_{\kb (t,t')}(t,t') = \frac{\langle \rho^{*}_{\kb (t,t')} e_-^{\int_{t'}^t ds\,\Omega^{\dagger}(s)} 
\rho_{\kb}^{}\rangle}{N S_{k(t,t')}}.
\end{eqnarray}
The definition in (\ref{correlator_def}) will prove useful later on (in Eq.(\ref{correlator})), 
when the advection of the wavevector is included into a single time evolution operator.
The influence of external flow enters (\ref{constit1}) both explicity, via the velocity 
gradient tensor in the prefactor, and implicitly through the reverse-advected wavevector. 
The transient correlator captures the slow structural relaxation of 
dense suspensions, thus rendering (\ref{constit1}) a useful tool for the study of arrested 
states. 

Before proceeding to develop a theory for the transient correlator, it is useful to 
first rearrange (\ref{constit1}) into an alternative form. 
To this end we introduce the Finger tensor \cite{doi}, a standard deformation measure 
from elasticity theory 
\begin{eqnarray}
\B(t,t')=\E(t,t')\E^T(t,t'),
\end{eqnarray} 
where the deformation tensor is given by (\ref{def}). 
The Finger tensor describes the stretching (but not rotation) of vectors embedded in the material. 
If we consider the quadratic form $\kb\!\cdot\!\B(t,t')\!\cdot\!\kb$ and take a derivative with 
respect to $t'$, we obtain
\begin{eqnarray}
\hspace*{-0.5cm}\partial_{t'}\,\big(\kb\!\cdot\!\B(t,t')\!\cdot\!\kb\big)&=&
%\kb\cdot\!\big(\,
%\E(t,t')\kap(t')\E^T(t,t') 
%\\ 
%&&\hspace*{0.7cm}\E(t,t')\kap^T(t')\E^T(t,t')\big)\!\cdot\kb
%\notag\\
\kb(t,t')\cdot\!\big(\, \kap(t) + \kap^T(t') \big)\!\cdot\kb(t,t')
\notag\\
&=&2\,\kb(t,t')\cdot\kap(t)\cdot\kb(t,t'),
\label{B_kap}
\end{eqnarray}  
where we have used the definition of the velocity gradient (\ref{velocity_gradient}).
Using (\ref{B_kap}) we can express our constitutive equation in the following alternative form  
\begin{eqnarray}
\hspace*{-2cm}\sig(t) &=& -\int_{-\infty}^{t} \!\!\!\!\!\!dt'\!\int\!\!\!
\frac{d{\bf k}}{32\pi^3} \left(\frac{\partial}{\partial t'}(\kb\!\cdot\!\B(t,t')\!\cdot\!\kb)\right)
\times
\notag\\
&&\hspace*{1cm}\times
\frac{\kb\kb}{k}\!
\left(\!\frac{S'_k S'_{k(t,t')}}{k(t,t')S^2_k}\!\right)\Phi_{{\bf k}(t,t')}^2(t,t').
\label{nonlinear}
\end{eqnarray}    
The expression (\ref{nonlinear}) provided the starting point for the development of a simpler 
`schematic' constitutive equation in \cite{pnas}. 

From (\ref{nonlinear}) it can be seen that the stress tensor at any given time is built by 
summing the contributions from all possible Fourier modes.  
Each of these individual contributions is obtained by integrating up the strain measure 
$\B(t,t')$ over the entire deformation history, weighted by a decaying, wavevector-dependent 
memory function. 
It is thus clear that strains occurring 
much further in the past than the longest relaxation time of $\Phi_{\kb}(t,t')$ cannot contribute to 
the integral; such distant strain increments have been forgotten. 
Moreover, physical intuition suggests that stress contributions at differing wavevector 
should be coupled, as the physical mechanisms driving relaxation on different length-scales 
are not independent (a fact ignored in many empirical constitutive equations).  
Within the mode-coupling approximations to be developed below this coupling becomes 
manifest in the nonlinear functional dependence of the memory function on the transient 
correlator; the equation of motion for $\Phi_{{\kb}}(t,t')$ is nonlocal in $\kb$-space.

\section{Equation of Motion}\label{equation_of_motion}
In order to close our constitutive theory we require an equation of motion 
for the transient density correlator. 
The fundamental approximations to be applied are very much in the spirit of standard quiescent 
mode-coupling theory, however, the presence of external flow necessitates considerable preparatory 
manipulation of formal expressions, not required in quiescent mode-coupling, before the theory 
arrives in a form convenient for making closure approximations. 
The first step on this path is to identify the operators which advect the wavevector of a 
density fluctuation (either forwards or backwards) and thus capture the purely affine component 
of the particle motion.

\subsection{Advection operators}
It is a straightforward exercise to show that a density fluctuation at advected 
wavevector $\bar{{\bf k}}(t,t')$ can be generated from one at $\kb$ using the 
exponential advection operator
\begin{eqnarray}\label{advection_op}
e_+^{-\!\int_{t'}^t\!ds\,\delta\Omega^{\dagger}(s)}|\,\rho_{\kb}\rangle
=|\,\rho_{\bar{\kb}(t,t')}\rangle\,, 
\end{eqnarray}
where the flow term in the adjoint Smoluchowski operator is given by 
\begin{eqnarray}\label{delta_omega_dagger}
\delta\Omega^{\dagger}(t)=\sum_i \rb_i\cdot\kap^T(t)\cdot\bp_i. 
\end{eqnarray}
The advection operator describes the purely affine time-evolution of fluctuations 
which would occur in the absence of both Brownian motion and interparticle interactions. 
It is also useful to identify an anlogous advection operator which acts on a density 
fluctuation to its left
\begin{eqnarray}\label{advection_op_left}
\langle\,\rho^*_{\kb}|\,
e_-^{\!\int_{t'}^t\!ds\,{\overline{\delta\Omega}^{\dagger}(s)}}
=\langle\,\rho^*_{\bar{\kb}(t,t')}|\,, 
\end{eqnarray} 
where the operator in the exponent is given by 
\begin{eqnarray}\label{operator_bar}
\overline{\delta\Omega^{\dagger}}(t)=\sum_i \rb_i\cdot\kap^T(t)\cdot(\bp_i+{\bf F}_i). 
\end{eqnarray}
The force term ${\bf F}_i$ in (\ref{operator_bar}) arises as a consequence of  
the equilibrium distribution function in our definition of the scalar product (\ref{braket}). 
The operations inverse to (\ref{advection_op}) and (\ref{advection_op_left}) which 
generate fluctuations at reverse-advected wavevectors are given by
\begin{eqnarray}
e_-^{\!\int_{t'}^t\!ds\,\delta\Omega^{\dagger}(s)}|\,\rho_{\kb}\rangle
&=&|\,\rho_{\kb(t,t')}\rangle\,,
\\
\langle\,\rho^*_{\kb}|\,
e_+^{-\!\int_{t'}^t\!ds\,{\overline{\delta\Omega}^{\dagger}(s)}}
&=&\langle\,\rho^*_{\kb(t,t')}|\,, 
\end{eqnarray} 
as can be determined from the inverse relations (\ref{unity}).

\subsection{Generalized diffusion kernel}
Using (\ref{advection_op}) the transient density correlator (\ref{correlator_def}) can 
be rewritten as  
\begin{eqnarray}\label{correlator}
\Phi_{\bf k}(t,t')&=&
\frac{1}{N S_{k}}
\langle\,\rho^*_{\bf k}
\,U(t,t')
\rho_{{\kb}} \rangle, 
\end{eqnarray}
where we have defined the propagator $U(t,t')$ as a product of 
two time-evolution operators
\begin{eqnarray}\label{prop1}
U(t,t')\equiv e_{-}^{\int_{t'}^t ds\,\Omega^{\dagger}(s)}
e_{+}^{-\int_{t'}^t ds\,\delta\Omega^{\dagger}(s)}.
\end{eqnarray}
In order to better understand the properties of $U(t,t')$ it is useful 
to express it in terms of a single exponential. 
Differentiation of (\ref{prop1}) with respect to $t$ and using (\ref{unity}) yields
\begin{eqnarray}
\frac{\partial}{\partial t}U(t,t') &=& U(t,t')\left[ e_{-}^{\int_{t'}^t ds\,\delta\Omega^{\dagger}(s)}
\,\Omega_e^{\dagger}\,
e_{+}^{-\int_{t'}^t ds\,\delta\Omega^{\dagger}(s)}
\right]
\notag\\
&\equiv& U(t,t')\Omega_{\rm nh}^{\dagger}(t,t'),
\label{prop1_eom}
\end{eqnarray}
where the subscript on $\Omega_{\rm nh}^{\dagger}(t,t')$ indicates that this operator is non-Hermitian.  
The formal solution of (\ref{prop1_eom}) is given by 
\begin{eqnarray}\label{nonherm}
U(t,t')=e_-^{\int_{t'}^t ds \,\Omega_{\rm nh}^{\dagger}(s,t')}.
\end{eqnarray}

Our aim to express the transient correlator (\ref{correlator}) in a form suitable for 
mode-coupling approximation is something which can only be achieved with a certain degree of foresight, 
combined with experience regarding the structure of the quiescent mode-coupling theory. 
This has the unavoidable consequence that the motivation behind some of our formal manipulations 
can only be fully appreciated in retrospect. 
We will attempt to highlight these potentially obscure steps when they occur and to motivate them 
as effectively as possible.  

Within the quiescent mode-coupling theory the Hermiticity of the equilibrium Smoluchowski 
operator leads to a positive semi-definite `initial decay rate' 
$\Gamma_{\kb}\equiv-\langle\rho^*_{\kb} \Omega_e^{\dagger} \rho_{\kb}\rangle/ NS_k$
determining the short-time dynamics of the transient correlator (see Eq.(\ref{first_eom}) below). 
Unfortunately, the operator $\Omega_{\rm nh}^{\dagger}(t,t')$ appearing in 
(\ref{nonherm}) does not retain this desirable feature and, if not treated carefully, can 
lead to a time-dependent initial decay rate $\Gamma_{\kb}(t,t')$ which changes sign as a 
function of accumulated strain. 
The resulting zeros of $\Gamma_{\kb}(t,t')$ cause divergences in the memory function 
$m_{\kb}(t,s,t')$ (see the denominator of (\ref{memory_friction})) which  
destroy the numerical stability of the theory. 
These considerations motivate us to seek a reformulation of the propagator (\ref{prop1}) 
for which the Hermitian operator 
\begin{eqnarray}\label{omega_a}
\Omega_a^{\dagger}(t,t')=e_{-}^{\int_{t'}^t ds\,\overline{\delta\Omega^{\dagger}}(s)}
\,\Omega_e^{\dagger}\,
e_{+}^{-\int_{t'}^t ds\,\delta\Omega^{\dagger}(s)},
\end{eqnarray}
becomes responsible for generating the initial 
decay rate of the theory. 
The operator (\ref{omega_a}) possesses the appealing feature that the average 
$\langle\rho^*_{\kb} \Omega_a^{\dagger}(t,t') \rho_{\kb}\rangle$ has the same form as that obtained 
from the equilibrium adjoint Smoluchowski operator, except that wavevectors are replaced by their 
advected counterparts. 
Positivity of the initial decay rate is thus ensured from the outset.

The operator $\Omega_a^{\dagger}(t,t')$ can be brought into the game by 
defining the time-dependent projection operator $P(t,t')$ and complement $Q(t,t')$
\begin{eqnarray}\label{time_projector}
P(t,t')&=&\sum_{\kb} |\rho_{\bar{\kb}(t,t')}\rangle\frac{1}{N S_{\bar{k}(t,t')}}\langle \rho^*_{\bar{\kb}(t,t')}|
\\
Q(t,t')&=&1-P(t,t'),
\label{11projector}
\end{eqnarray}
and employing the following exact operator identity to rewrite the propagator 
\begin{eqnarray}\label{big_identity}
&&\hspace*{-1cm}e_-^{\int_{t'}^t ds \,\Omega_{\rm nh}^{\dagger}(s,t')}
=
e_-^{\int_{t'}^t ds \,\mathcal{G}(s,t')} 
\\
&&\hspace*{-1.cm}+\int_{t'}^{t}\!\!ds\, e_-^{\int_{t'}^sds' \Omega_{\rm nh}^{\dagger}(s',t')}
\mathcal{H}(s,t')\,
e_-^{\int_{s}^{t}ds' \mathcal{G}(s',t')},
\notag
\end{eqnarray}
where the operators $\mathcal{G}(t,t')$ and $\mathcal{H}(t,t')$ are given by 
\begin{eqnarray}\label{g_operator}
\mathcal{G}(t,t')&=&e_{-}^{\int_{t'}^t ds\,\delta\Omega^{\dagger}(s)}
\,Q(t,t')\,\Omega_e^{\dagger}\,
e_{+}^{-\int_{t'}^t ds\,\delta\Omega^{\dagger}(s)}
\\
\mathcal{H}(t,t')&=&e_{-}^{\int_{t'}^t ds\,\delta\Omega^{\dagger}(s)}
\,P(t,t')\,\Omega_e^{\dagger}\,
e_{+}^{-\int_{t'}^t ds\,\delta\Omega^{\dagger}(s)}.
\end{eqnarray}
The identity (\ref{big_identity}) may be verified by differentiation and, 
despite its complex structure, is simply a further time-dependent generalization of 
(\ref{zwanzig_identity}) employed earlier. 
%It has been constructed such as to generate an initial decay rate which remains positive 
%definite under flow and which reduces to the equilibrium quantity for quiescent states. 
Note that the advection operator $\overline{\delta\Omega^{\dagger}}(s)$ which 
is needed to build our desired Hermitian operator $\Omega_a^{\dagger}(t,t')$ enters the calculation 
via the advected density fluctuation on the right of the projector (\ref{time_projector}).

We next take the derivative of (\ref{big_identity}) with respect to $t$ and use the lengthy expression 
obtained to calculate matrix elements between density fluctuations 
$\langle\,\rho^*_{\kb}|\cdots|\rho^{}_{\kb}\rangle/NS_k$. 
This calculation yields an equation of motion for the 
transient density correlator featuring a generalized diffusion kernel
\begin{eqnarray}\label{first_eom}
\frac{\partial}{\partial t}\Phi_{\bf k}(t,t') &+& 
\Gamma_{\bf k}(t,t')\Phi_{\bf k}(t,t') \\
&+& \int_{t'}^{t}\!ds \,M_{\bf k}(t,s,t')\Phi_{\bf k}(s,t') 
=\Delta_{\bf k}(t,t').
\notag
\end{eqnarray}
Because matrix elements are taken with respect to a density fluctuation at a single wavevector,  
translational invariance eliminates the wavector sum appearing in (\ref{time_projector}). 

The initial decay rate $\Gamma_{\bf k}(t,t')$ depends upon the Hermitian operator 
$\Omega_a^{\dagger}(t,t')$ and can be explicitly evaluated in terms of the advected wavevector 
and static structure factor 
\begin{eqnarray}\label{initial_decay}
\Gamma_{\bf k}(t,t') = 
-\frac{\langle\rho^*_{\kb} \Omega_a^{\dagger}(t,t') \rho^{}_{\kb}\rangle}{NS_{\bar{k}(t,t')}}
=
\frac{\bar{k}^2(t,t')}{S_{\bar{k}(t,t')}}.
\end{eqnarray}
The initial decay rate takes the same functional form as in equilibrium, albeit with the static wavevector 
replaced by the advected one. 
For noninteracting particles only the first two terms in (\ref{first_eom}) survive 
and we obtain the exact solution for the correlator in this case, incorporating the phenomena 
of Taylor dispersion (enhanced diffusion in the direction of flow) \cite{foister}.  

The generalized diffusion kernel entering (\ref{first_eom}) is given by the formal expression 
\begin{eqnarray}\label{diffusion_kernel}
M_{\bf k}(t,s,t')=
-\frac{\langle  B_{\kb}^*(s,t') \,\tilde{U}(t,s,t') A^{}_{\kb}(t,t') \rangle}{NS_{\bar{k}(s,t')}}
\end{eqnarray}
where we have defined several new operators in an effort to streamline the notation
\begin{eqnarray} 
%Q_a(t,t')&=&e_{-}^{\int_{t'}^t ds\,\delta\Omega^{\dagger}(s)}
%\,Q(t,t')\,
%e_{+}^{-\int_{t'}^t ds\,\overline{\delta\Omega^{\dagger}}(s)} 
%\\
\tilde{U}(t,s,t')&=&e_-^{\int_s^t \!ds' \mathcal{G}(s'\!,t') } 
\label{U_tilde}
\\
B_{\kb}^*(t,t')&=&\rho_{\kb}^{*}\Omega_a^{\dagger}(t,t')
\label{operatorB}
\\
A^{}_{\kb}(t,t')&=&\mathcal{G}(t,t')\rho_{\kb}^{}.
%Q_a(t,t')\Omega_a^{\dagger}(t,t')\rho_{\kb}^{}
\label{operatorA}
\end{eqnarray}
A striking feature of (\ref{diffusion_kernel}) is that the diffusion kernel is a function of three 
time arguments. 
We recall that for quiescent systems the memory function depends upon only a time 
difference \cite{goetze}, whereas for systems under steady shear it depends upon two time 
arguments \cite{fc09}. 
The former property is a consequence of the time translational invariance of the quiescent 
state (ageing of arrested states being neglected within MCT), whereas the latter property is formal recognition of  
the existence of an absolute reference time, namely the time at which the flow 
was switched on. In practice, however, the residual transient may be vanishingly small 
for the steady states of interest. 
The three-time argument memory function (\ref{diffusion_kernel}) may be interpreted as describing 
the decay of memory between times $s$ and $t$, incorporating the coupling to the stress which is still relaxing from the strain accumulated between times $t'$ and $t$.

The final term in the equation of motion for the transient correlator (\ref{first_eom}) is 
the `remainder' term on the right hand side, given by
\begin{eqnarray}
\label{remainder}
\Delta_{\bf k}(t,t')=\frac{1}{NS_k}\langle\rho^*_{\kb}
\tilde{U}(t,t',t') \mathcal{G}(t,t') 
\rho^{}_{\kb}\rangle. 
\end{eqnarray}
We know rather little about the properties of this term and the fact that it 
remains finite represents a primary drawback of the 
the present approach. (Below we will need to approximate it by zero.)
Its existence stems from our use of the time-dependent projection operator 
(\ref{11projector}) which only eliminates linear coupling to the density at $t=t'$, leading to 
$\Delta_{\bf k}(t,t)=0$, but not at differing values of the time arguments.

\subsection{Generalized friction kernel}
Eq.(\ref{first_eom}) represents an exact equation of motion for the transient density correlator 
in terms of a generalized diffusion kernel. However, experience with quiescent mode-coupling 
theory has shown that it is difficult to accurately approximate a diffusion kernel and that far 
better results can be obtained when starting from an equation of motion involving a friction kernel 
\cite{fuchs2005}.   
We thus seek to convert the generalized diffusion kernel $M_{\bf k}(t,s,t')$ to a generalized 
friction kernel $m_{\bf k}(t,s,t')$.
Following standard mode-coupling arguments, the transformation from a diffusion to a friction 
kernel is achieved by splitting the time-evolution operator entering the diffusion kernel, $\tilde{U}(t,s,t')$ in the present case, into `reducible' and `irreducible' components. 
As with our previous formal manipulations, foresight is required in making 
this step. 
The specific choice of how to split the time-evolution operator is made strategically, with a 
view to obtain an equation of the form (\ref{Mm_equation}) which, together with (\ref{first_eom}) 
can then be conveniently solved to yield (\ref{second_eom}), the desired equation 
of motion. 

The first step is to take the time derivative of $\tilde{U}(t,s,t')$. 
Using (\ref{U_tilde}), this yields
\begin{eqnarray}
\frac{\partial}{\partial t}\tilde{U}(t,s,t')=\tilde{U}(t,s,t')\mathcal{G}(t,t').
\end{eqnarray}
We next define a non-Hermitian, time-dependent projection operator 
\begin{eqnarray}\label{P_tilde}
\tilde{P}(t,t')=\sum_{\kb}|\,\rho^{}_{\kb}\rangle\frac{1}{\langle\rho^{*}_{\kb} \Omega_a^{\dagger}(t,t') \rho^{}_{\kb}\rangle} \langle\rho^{*}_{\kb}
\Omega_a^{\dagger}(t,t')|,
\end{eqnarray}
with complement $\tilde{Q}(t,t')=1-\tilde{P}(t,t')$, and use this to split $\mathcal{G}(t,t')$ into two 
contributions
\begin{eqnarray}\label{split_differential}
\frac{\partial}{\partial t}\tilde{U}(t,s,t')=&&\!\!\!\!\!\!\tilde{U}(t,s,t')\mathcal{G}(t,t')\big(\tilde{Q}(t,t')
+\tilde{P}(t,t')\big)
\\
=&&\!\!\!\!\!\!\tilde{U}(t,s,t')\mathcal{G}^{\rm irr}(t,t') + \tilde{U}(t,s,t')\mathcal{G}^{\rm red}(t,t').
\notag
\end{eqnarray}
The irreducible operator is simply defined by $\mathcal{G}^{\rm irr}(t,t')=\mathcal{G}(t,t')\tilde{Q}(t,t')$,  
nothing more can be done with it, whereas the reducible operator can be expressed in terms of 
previously encountered quantities 
\begin{eqnarray}\label{reducible}
\mathcal{G}^{\rm red}(t,t')= |A^{}_{\kb}(t,t')\rangle\frac{1}{S_k\Gamma_{\kb}(t,t')}\langle B^{*}_{\kb}(t,t')|
\end{eqnarray}
Treating the term $\tilde{U}(t,s,t')\mathcal{G}^{\rm red}(t,t')$ appearing in (\ref{split_differential}) 
as an inhomogeneity leads to the formal solution 
\begin{eqnarray}\label{red_irr}
\tilde{U}(t,s,t') &=& U^{\rm irr}(t,s,t') \\
&+& \int_s^t \!dt''\,\tilde{U}(t''\!,s,t')\mathcal{G}^{\rm red}(t''\!,t')
U^{\rm irr}(t,t''\!,t'),
\notag
\end{eqnarray}
where the irreducible time evolution is given by  
\begin{eqnarray}
U^{\rm irr}(t,s,t')=e_-^{\int_{s}^{t}ds'\mathcal{G}^{\rm irr}(s',t')}, 
\end{eqnarray}
which is the solution of the corresponding homogeneous equation.
Using (\ref{operatorB}) and (\ref{operatorA}) to take matrix elements of (\ref{red_irr}), 
$\langle B^{*}_{\kb}(s',t')|\cdots| A^{}_{\kb}(t,t') \rangle$, and using (\ref{reducible}) thus 
leads to an important equation relating the diffusion and friction kernels
\begin{eqnarray}\label{Mm_equation}
M_{\kb}(t,s,t') &=& -\Gamma_{\kb}(s,t')m_{\kb}(t,s,t')\Gamma_{\kb}(t,t') \\
&-& \int_s^t \!\!dt''\, M_{\kb}(t'',s,t')m_{\kb}(t,t'',t')\Gamma_{\kb}(t,t').\notag
\end{eqnarray}
The generalized friction kernel is given by 
\begin{eqnarray}\label{memory_friction}
\!m_{\bf k}(t,s,t')
\!= \frac{
\langle
B^{*}_{\kb}(s,t') U^{\rm irr}(t,s,t')
A^{}_{\kb}(t,t') 
\rangle}
{
NS_{\bar{k}(s,t')} \Gamma_{\bf k}(s,t') \Gamma_{\bf k}(t,t')
}\,, 
\end{eqnarray}
which involves the irreducible dynamics. 
At this stage in the calculation the importance of enforcing a positive definite 
initial decay rate becomes clearer. 
Alternative formulations for which advection can cause $\Gamma_{\bf k}(t,t')$ to change 
sign for some values of $t$ and $t'$ will inevitably lead to undesirable singularities in the memory 
kernel (\ref{memory_friction}).
While such singularities could, in principle, be integrable and thus remain physical, they 
would at the very least prove inconvenient for numerical implementations of the 
theory.

We have now almost arrived at our goal to reformulate the equation of motion. 
The final step is to combine (\ref{first_eom}) with (\ref{Mm_equation}), with the former 
expressed as an inhomogeneous integral equation
\begin{eqnarray}\label{inhom}
\Phi_{\bf k}(t,t')
=\frac{-1}{\Gamma_{\bf k}(t,t')}\int_{t'}^{t}\!ds \,M_{\bf k}(t,s,t')\Phi_{\bf k}(s,t') 
+g_{\kb}(t,t').
\notag\\
\end{eqnarray}
This equation can be viewed as a Volterra equation of the second kind \cite{tricomi} with 
an inhomogeneity given by 
\begin{eqnarray}
g_{\kb}(t,t')=
\frac{-1}{\Gamma_{\bf k}(t,t')}\left( \frac{\partial}{\partial t}\Phi_{\bf k}(t,t') 
-\Delta_{\bf k}(t,t')
\right). 
\end{eqnarray}
The solution of this equation follows directly from the theory of Volterra equations 
(see Appendix \ref{volterra}) and yields the final form for the equation of 
motion
\begin{eqnarray}\label{second_eom}
\hspace*{0.cm}
\frac{\partial}{\partial t}\Phi_{\bf k}(t,t')
&+& \Gamma_{\bf k}(t,t')\bigg(
\Phi_{\kb}(t,t')
\\
&&\hspace*{-1cm}+
\int_{t'}^t ds \,m_{\kb}(t,s,t') \frac{\partial}{\partial s} \Phi_{\kb}(s,t')
\bigg) = \tilde\Delta_{\bf k}(t,t'),
\notag
\end{eqnarray} 
where the modified remainder term $\tilde\Delta_{\bf k}(t,t')$ is given in terms of previously defined 
quantities 
\begin{eqnarray}\label{delta_integral}
\tilde\Delta_{\bf k}(t,t') = \Delta_{\bf k}(t,t') 
+ \Gamma_{\bf k}(t,t')\int_{t'}^{t}\!\!ds \,m_{\bf k}(t,s,t')\Delta_{\bf k}(s,t'). 
\notag\\
\end{eqnarray}
Although the integral expression (\ref{delta_integral}) appears to be rather intractable, 
it can be simplified to a single equilibrium average. 
Substituting the formal solution (\ref{red_irr}) into our expression for the remainder term 
(\ref{remainder}) 
recovers (\ref{delta_integral}) with the modified remainder identified as
\begin{eqnarray}
\label{explicit_delta}
\tilde\Delta_{\bf k}(t,t')
=
\frac{1}{NS_k}\langle
\rho^{*}_{\kb}U^{\rm irr}(t,t',t')\mathcal{G}(t,t')\rho^{}_{\kb}
\rangle, 
\end{eqnarray}
by inspection. 
The time evolution of $\tilde\Delta_{\bf k}(t,t')$ appearing in the equation of motion for the correlator 
(\ref{second_eom}) is thus shown to be determined by the irreducible dynamics.

\subsection{Closure approximation}
The equation of motion (\ref{second_eom}) we have obtained is formally exact, but still contains two unknown 
quantities, the generalized friction kernel $m_{\kb}(t,s,t')$ and the remainder term 
$\tilde\Delta_{\bf k}(t,t')$, both of which require approximation to arrive at a closed theory. 
This final stage in our development of a constitutive theory consists of two distinct steps. 
The first of these is specific to systems under external flow and involves the neglect of certain `strain 
energy' terms, which, it is hoped, are of minor importance in determining the relaxation of the transient 
correlator. 
Our choice to neglect these terms, which do not arise in quiescent mode-coupling theory, represents the 
only point in our development at which an approximation is made which lies beyond the established canon of 
mode-coupling projections and Gaussian factorizations. 
Once accepted, this new approximation yields directly an appealingly symmetric form for the friction 
kernel (\ref{memory_friction}) and causes the remainder term (\ref{explicit_delta}) to vanish 
identically. 
The second and final step is then a rather standard projection of the friction kernel onto density pairs 
and factorization of the resulting four-point average. 

We consider first the remainder term (\ref{explicit_delta}). 
As described in detail in Appendix \ref{simplifying}, making the assumption that the  
two strain-energy terms introduced there
\begin{eqnarray}\label{big_sigma1}
&&\hspace*{-0.9cm}
\Sigma(t,t') \!\equiv\!\!
\int_{t'}^t\!\!ds \, e_+^{-\int_{s}^t \!ds'\, \overline{\delta\Omega^{\dagger}}(s')}
\kap^{T}(s)\!:\!\hat{\sig}\,e_-^{\int_{s}^t \!ds'\, \delta\Omega^{\dagger}(s')} 
\approx 0
\notag
\\
\\
\label{big_sigma2}
&&\hspace*{-0.9cm}
\overline{\Sigma}(t,t') \!\equiv\!\!
\int_{t'}^t\!\!ds \, e_+^{-\int_{s}^t \!ds'\, \delta\Omega^{\dagger}(s')}
\kap^{T}(s)\!:\!\hat{\sig}\,e_-^{\int_{s}^t \!ds'\, \overline{\delta\Omega^{\dagger}}(s')}
\approx 0,
\notag
\\
\end{eqnarray}
both vanish, gives rise to the simplifications 
\begin{eqnarray}
\mathcal{G}(t,t')&\approx& \mathcal{G}_Q(t,t')
\\
\label{approx2}
U^{\rm irr}(t,s,t')
&\approx&
U^{\rm irr}_Q(t,s,t'), 
\end{eqnarray}
where $\mathcal{G}_Q(t,t')$ and $U^{\rm irr}_Q(t,s,t')$  are given by (\ref{GQ}) and (\ref{perpendicular_dynamics}), 
respectively. 
We thus find that the remainder term vanishes identically 
\begin{eqnarray}
\tilde\Delta_{\bf k}(t,t')
\approx
\frac{1}{NS_k}\langle
\rho^{*}_{\kb}U_Q^{\rm irr}(t,t',t')\mathcal{G}_Q(t,t')\rho^{}_{\kb}
\rangle=0,
\end{eqnarray}
as both operators in the average are orthogonal to linear density fluctuations.

Our assumption that the strain-energy terms (\ref{big_sigma1},\ref{big_sigma2}) 
vanish is certainly an uncontrolled approximation, but 
may be partially motivated by considering their dependence upon strain. For the case of steady 
flow \cite{fc09}, 
both $\Sigma$ terms increase in proportion to the strain accumulated since startup and thus have 
an influence which depends upon whether the correlator has relaxed completely for small strain values. 
For glassy states our neglect of the $\Sigma$ terms is thus equivalent to requiring that the yield 
strain remain small, which is consistent with simulation results \cite{zausch,miyazaki2004,varnik}. 
In any case, if the yield strain were found to be large, then the implicit assumption of a harmonic 
free energy functional underpinnning the entire mode-coupling approach would be invalidated, such that 
the new strain-energy terms would be among the least of our worries.

We turn now to the friction kernel (\ref{memory_friction}).  
Using the previously derived relations (\ref{advection_op}), (\ref{advection_op_left}), 
(\ref{omega_a}), (\ref{g_operator}), (\ref{operatorB}) and (\ref{operatorA}) we write the 
numerator in the following, somewhat more explicit, form
\begin{eqnarray}
&&\hspace*{-1cm}\langle
B^{*}_{\kb}(s,t') U^{\rm irr}(t,s,t')
A^{}_{\kb}(t,t') 
\rangle=
\\
&&\hspace*{-1cm}\langle 
\rho_{\bar{\kb}(s,t')}^{*}\Omega_e^{\dagger}e_{+}^{-\int_{t'}^s ds'\,\delta\Omega^{\dagger}(s')}
U^{\rm irr}(t,s,t')\times
\notag\\
&&\hspace*{1cm}\times
e_{-}^{\int_{t'}^t ds\,\delta\Omega^{\dagger}(s)}
Q(t,t')\Omega_e^{\dagger} \rho_{\bar{\kb}(t,t')}^{}
\rangle.
\notag
\end{eqnarray}
Approximation (\ref{big_sigma2}) enables us to replace $\delta\Omega^{\dagger}(s)$ in 
the leftmost advection operator by $\overline{\delta\Omega^{\dagger}}(s)$ (see (\ref{replacement2}))

\begin{eqnarray}\label{approx1}
e_+^{-\int_{t'}^s \!ds'\, \delta\Omega^{\dagger}(s')} 
&\approx&
e_+^{-\int_{t'}^s \!ds'\, \overline{\delta\Omega^{\dagger}}(s')},  
\end{eqnarray}
which, together with (\ref{approx2}), can be used to cast the numerator of the 
friction kernel, Eq.(\ref{memory_friction}), in the more symmetrical form
\begin{eqnarray}\label{symmetrized}
&&\!\!\!\!\langle 
\rho_{\bar{\kb}(s,t')}^{*}\Omega_e^{\dagger}
\,Q(s,t')\,
e_{+}^{-\int_{t'}^s ds'\,\overline{\delta\Omega^{\dagger}}(s')}
U_Q^{\rm irr}(t,s,t')
\times
\notag\\
&&\hspace*{1.3cm}\times
e_{-}^{\int_{t'}^t ds\,\delta\Omega^{\dagger}(s)}
Q(t,t')\,\Omega_e^{\dagger}\, \rho_{\bar{\kb}(t,t')}^{}
\rangle.
\end{eqnarray}
The projected irreducible propagator $U^{\rm irr}_Q(t,s,t')$ is given by 
(\ref{perpendicular_dynamics}) and acts in the space perpendicular to linear 
density fluctuations. 
It is this fact which enables us to insert, without incurring further approximation, the 
extra projector $Q(t,t')$ on the left of the irreducible propagator. 
The quantities $Q(t,t')\,\Omega_e^{\dagger}\, \rho_{\bar{\kb}(t,t')}$ appearing on either 
side of the propagator in (\ref{symmetrized}) are fluctuating forces which do not couple to 
linear density fluctuations.

We next introduce the time-dependent 
projection operator onto density pairs
\begin{eqnarray}\label{pair_projector_time}
P_2(t,t')=\sum_{\qb > \pb} 
\frac{
|\,\rho^{}_{\bar{\qb}(t,t')}\rho^{}_{\bar{\pb}(t,t')}\rangle 
\langle\rho^{*}_{\bar{\qb}(t,t')}\rho^{*}_{\bar{\pb}(t,t')}|
}{N^2 S_{\bar{q}(t,t')} S_{\bar{p}(t,t')}} ,
\end{eqnarray}
which is simply (\ref{pair_projector}) evaluated at the advected wavevector and  
therefore subject to the same Gaussian approximation in the denominator. 
The numerator (\ref{symmetrized}) is thus approximated by 
\begin{eqnarray}\label{symmetrized_approx}
&&\!\!\!\!\langle 
\rho_{\bar{\kb}(s,t')}^{*}\Omega_e^{\dagger}
\,Q(s,t')P_2(s,t')\,
e_{+}^{-\int_{t'}^s ds'\,\overline{\delta\Omega^{\dagger}}(s')}
U_Q^{\rm irr}(t,s,t')
\times
\notag\\
&&\hspace*{1cm}\times\;e_{-}^{\int_{t'}^t ds\,\delta\Omega^{\dagger}(s)}
P_2(t,t')Q(t,t')\,\Omega_e^{\dagger}\, \rho_{\bar{\kb}(t,t')}^{}
\rangle, 
\notag\\
\notag\\
&&\hspace*{-0.25cm}\approx\hspace*{-0.25cm}\mathop{\sum_{\qb > \pb}}_{\;\;\,\qb' > \pb'}
\hspace*{-0.2cm}
\frac{V^{(1)}_{\kb\qb\pb}(s,t')
V^{(2)}_{\kb\qb'\pb'}(t,t')}{N^2}
\langle\rho^{*}_{\bar{\qb}(s,t')}\rho^{*}_{\bar{\pb}(s,t')}
\times
\notag\\
&&\hspace*{-0.3cm}\times\,
e_{+}^{-\int_{t'}^s ds'\,\overline{\delta\Omega^{\dagger}}(s')}
U_Q^{\rm irr}(t,s,t')
\;e_{-}^{\int_{t'}^t ds\,\delta\Omega^{\dagger}(s)}
\rho^{}_{\bar{\qb}'(t,t')}\rho^{}_{\bar{\pb}'(t,t')}\rangle,
\notag\\
\label{numerator_fourpoint}
\end{eqnarray}
where the vertex functions are given by 
\begin{eqnarray}
&&\hspace*{-1cm}V^{(1)}_{\kb\qb\pb}(s,t') \!=\! 
\frac{\langle\rho_{\bar{\kb}(s,t')}^{*}\Omega_e^{\dagger}
\,Q(s,t')\rho^{}_{\bar{\qb}(s,t')}\rho^{}_{\bar{\pb}(s,t')}\rangle}
{N S_{\bar{q}(s,t')} S_{\bar{p}(s,t')}}
\\
&&\hspace*{-1cm}V^{(2)}_{\kb\qb'\pb'}(t,t') \!=\!
\frac{\langle
\rho^{*}_{\bar{\qb'}(t,t')}\rho^{*}_{\bar{\pb'}(t,t')}
Q(t,t')\,\Omega_e^{\dagger}\, \rho_{\bar{\kb}(t,t')}^{}
\rangle}
{N S_{\bar{q'}(t,t')} S_{\bar{p'}(t,t')}}.
\end{eqnarray}
Employing the factorization approximation for the triplet static structure factor 
$S^{(3)}_{\kb\qb\pb}\approx S_{\kb}S_{\qb}S_{\pb}$ enables the vertices to be expressed 
in terms of known two-point static correlations
\begin{eqnarray}
\label{v1}
&&\hspace*{-0.4cm}V^{(1)}_{\kb\qb\pb}(s,t') \!= \bar\kb(s,t')\!\cdot\!\big(
\bar\qb(s,t') c_{\bar{q}(s,t')} \!+ 
\bar\pb(s,t') c_{\bar{p}(s,t')}\big)
\times
\notag\\
&&\hspace*{4.cm}\times
\rho\, \delta_{\bar{\kb},\bar{\qb}+\bar{\pb}}
\\
\label{v2}
&&\hspace*{-0.4cm}V^{(2)}_{\kb\qb'\pb'}(t,t') \!= \bar\kb'(t,t')\!\cdot\!\big(
\bar\qb'(t,t') c_{\bar{q}'(t,t')} \!+ 
\bar\pb'(t,t') c_{\bar{p}'(t,t')}\big)
\!\times
\notag\\
&&\hspace*{4.cm}\times
\rho\, \delta_{\bar{\kb}',\bar{\qb}'+\bar{\pb}'},
\end{eqnarray}
where $\rho=N/V$ is the particle number density and $c_q=(1-1/S_q)/\rho$ is the 
Ornstein-Zernike direct correlation function. 
 The four-point correlator entering (\ref{numerator_fourpoint}) is approximated in the spirit 
of quiescent mode-coupling theory by factorizing and replacing the $Q$-projected irreducible 
dynamics by the full dynamics
\begin{eqnarray}
&&\hspace*{-0.7cm}\langle\rho^{*}_{\bar{\qb}(s,t')}\rho^{*}_{\bar{\pb}(s,t')}
e_{+}^{-\int_{t'}^s ds'\,\overline{\delta\Omega^{\dagger}}(s')}
U_Q^{\rm irr}(t,s,t')\times
\notag\\
&&
\hspace*{2cm}\times\; e_{-}^{\int_{t'}^t ds\,\delta\Omega^{\dagger}(s)}
\rho^{}_{\bar{\qb}'(t,t')}\rho^{}_{\bar{\pb}'(t,t')}\rangle
\notag\\
&&\hspace*{-0.85cm}
\approx
N^2 S_{\bar{q}(s,t')}S_{\bar{p}(s,t')}\Phi_{\bar{\kb}(s,t')}(t,s)
\Phi_{\bar{\pb}(s,t')}(t,s)
\delta_{\qb,\qb'}\delta_{\kb,\kb'}
\notag\\
\label{fourpoint}
\end{eqnarray}
where the Kronecker delta functions enforce translational invariance. 
Using (\ref{numerator_fourpoint}) together with (\ref{v1}), (\ref{v2}) and 
(\ref{fourpoint}) to approximate the numerator in (\ref{memory_friction}) 
and employing the explicit expression (\ref{initial_decay}) for the initial decay rate 
in the denominator thus yields our final mode-coupling approximation to the memory function
\begin{eqnarray}\label{final_memory}
m_{\bf k}(t,s,t') \!\!&=& \!\!
\frac{\rho}{16\pi^3} \!\!\int \!\! d\qb 
\frac{S_{\bar{k}(t,t')} S_{\bar{q}(s,t')} S_{\bar{p}(s,t')} }
{\bar{k}^2(s,t') \bar{k}^2(t,t')}\times
\\
&&\hspace*{-1.5cm}\times V_{\kb\qb\pb}(s,t')\,V_{\kb\qb\pb}(t,t')\Phi_{\bar{\qb}(s,t')}(t,s)
\Phi_{\bar{\pb}(s,t')}(t,s).
\notag
\end{eqnarray}
The wavevector restrictions enforced by the Kronecker delta functions appearing in both the vertices 
and the mode-coupling factorization of the four-point correlation function reduce the four-fold sum 
in (\ref{numerator_fourpoint}) to a single sum over $\qb$ (which in (\ref{final_memory}) has been 
replaced by an integral) and leads to a coupling of the two `internal' wavevectors $\qb$ and $\pb$ via 
the condition $\pb=\qb-\kb$.

We have thus arrived at a closed constitutive equation for the rheology of colloidal suspensions which 
requires only the number density $\rho$, the static structure factor $S_q$ and velocity gradient tensor 
$\kap(t)$ as input. 
In order to calculate the stress tensor one must first solve the nonlinear integro-differential equation 
(\ref{second_eom}) for the transient density correlator using the approximation $\tilde{\Delta}_{\bf k}(t,t')=0$ 
and the approximate memory function (\ref{final_memory}) with vertex functions (\ref{v1}) and (\ref{v2}). 
The choice of flow enters via the advected wavevectors. 
The transient correlator thus obtained is then substituted into the constitutive equation (\ref{nonlinear}) 
and integrated. 
There are no adjustable parameters in our theory as such, although we accept that the approximations we 
have made are not the only ones possible. (In that sense, our theory is one of a larger family that could 
be specified by adjusting parameters that remain, at this stage, unidentified.)
In the absence of flow the theory for the transient correlator reduces to the familiar quiescent 
mode-coupling theory \cite{goetze}.

\subsection{Linear response}
It is informative to consider the slow-flow limit of the constitutive equation (\ref{nonlinear}) 
for which the stress response becomes linear in the velocity gradient tensor. 
To obtain the linear response expression we first neglect the strain dependence of the transient 
correlator and replace advected wavevectors by their non-advected counterparts, wherever these 
appear explicitly. We thus obtain  
\begin{eqnarray}
\hspace*{0cm}\sig(t) = -\!\int_{-\infty}^{t} \!\!\!\!\!\!\!\!dt'\!\int\!
\frac{\kb\kb \,d{\bf k}}{32\pi^3} \,\kb\!\cdot\!\frac{\partial \B(t,t')}{\partial t'}\!\cdot\!\kb
\!
\left(\!\frac{S'_k \Phi_{k}^2(t-t')}{k S_k} \!\right)^2\!\!\!\!,
\notag\\
\label{partial_linear}
\end{eqnarray} 
where the quiescent correlator depends only on a time difference. 
We next expand the Finger tensor to linear order in the velocity gradient tensor
\begin{eqnarray}
\B(t,t')&=&e_+^{\int_{t'}^t \!ds\, \kap(s)}e_-^{\int_{t'}^t \!ds\, \kap^T(s)}
\notag\\
&\approx& {\bf 1} + 2\int_{t'}^t\!ds\,\overline{\kap}(s),
\label{finger_expansion}
\end{eqnarray} 
where $\overline{\kap}=(\kap+\kap^T)/2$ is the symmetrized velocity gradient tensor.
Substitution of (\ref{finger_expansion}) into (\ref{partial_linear}) thus yields the 
desired linear response result
\begin{eqnarray}
\hspace*{-0.2cm}
\sig^{\rm l}(t)\!=\!\!\!\int_{-\infty}^t \!\!\!\!\!dt'\!\int\! d\kb
\{\left(\kb\!\cdot\!\overline{\kap}(t')\!\cdot\!\kb\right)\kb\kb\}
\left( \frac{S'_k\Phi_k(t-t')}{16\pi^3kS_k} 
\!
\right)
\label{linear}
\end{eqnarray}
The required quiescent correlator is obtained from solution of the $\kap(t)\rightarrow 0$ limit of the 
equation of motion (\ref{second_eom}) 
\begin{eqnarray}\label{quiescent_mct}
\hspace*{0.cm}
\frac{\partial}{\partial t}\Phi_k(t)
&+& \Gamma_k\bigg(
\Phi_k(t) +
\int_{0}^t \!ds \,m_k(t-s) \frac{\partial}{\partial s} \Phi_k(s)
\bigg) = 0,
\notag
\end{eqnarray} 
with initial decay rate $\Gamma_k=k^2/S_k$ and memory function 
\begin{eqnarray}\label{quiescent_memory}
m_k(t) \!&=& \!
\rho \!\!\int \!\! d\qb\, 
\frac{S_{k} S_{q} S_{p} }
{16\pi^3k^4}
V^2_{\kb\qb\pb}\Phi_{q}(t)
\Phi_{p}(t),
\end{eqnarray} 
where the vertex function is given by $V_{\kb\qb\pb} = \kb\cdot(\qb c_{q} + \pb c_{p} )$. 

The anisotropic part of the integrand appearing in (\ref{linear}) is contained within the factor 
$\{\hspace*{0.02mm}\cdot\}$. The angular $\kb$-integrals can therefore easily be performed for a given $\kap(t)$ and 
it follows that for a given steady flow rate $\dot{\gamma}$ the Trouton ratio 
of extensional viscosity $\eta_e\equiv (\sigma_{xx}-\sigma_{yy})/\dot{\gamma}$ to shear viscosity 
$\eta_s\equiv\sigma_{xy}/\dot{\gamma}$ takes the values $4$ and $3$ when the extensional flow is 
planar and uniaxial, respectively. 
These simple  geometrical ratios are in compliance with {\em Trouton's rules}, familiar to 
continuum rheologists. 
 
The elastic limit can be accessed by partial integration of (\ref{linear}) and yields 
Hooke's law for an incompressible isotropic elastic body
\begin{eqnarray}
\sig(t) = 2G_{\infty}\boldsymbol{\epsilon}(t).
\end{eqnarray}  
The infinitesimal accumulated strain is given by 
\begin{eqnarray}
\boldsymbol{\epsilon}(t)=\int_{-\infty}^t \!\!\!\!\!ds\;\overline{\kap}(s), 
\end{eqnarray} 
and the single elastic constant predicted by the theory is given by the standard mode-coupling 
expression for the plateau value of the shear modulus 
\cite{bergenholtz_naegele}
\begin{eqnarray} 
G_{\infty}=\frac{1}{60\pi^2}\int \!d\kb\;k^4 \left(\frac{S'_k}{S_k}\right)^2\!\!\Phi^2_k(t\to \infty).
\end{eqnarray} 
For arrested states we recall that the long time limit of the correlator $\Phi_k(t\to\infty)$ 
remains finite and serves as an order parameter for the transition from a fluid to 
a glass or gel. 
We note that the present theory makes no prediction for the bulk modulus, as the assumed 
compressible flow satisfies ${\rm Tr}\,\boldsymbol{\varepsilon}=0$.

\subsection{Material objectivity} 
The `principle of material objectivity' expresses the requirement that the constitutive relationship 
between stress and strain tensors should be invariant with respect to rotation of either the material 
body or the observer, thus preventing an unphysical dependence of the stress upon the state of rotation. 
Material objectivity is an approximate symmetry, based on the neglect of inertial effects on 
the microscopic level (i.e. the influence of centrifugal and coriolis forces on 
particle trajectories). 
Nevertheless, many soft materials display this symmetry to an excellent level of approximation. 
The overdamped Smoluchowski dynamics underlying our treatment excludes inertial 
effects from the outset and, providing that our approximations preserve this, our 
set of equations (\ref{nonlinear}),(\ref{second_eom}) and (\ref{final_memory}) should be material 
objective. 

Material objectivity can be explicity confirmed by using (\ref{advection2}) and (\ref{advection1}) to 
eliminate the advected wavevectors in favor of the deformation tensors $\E(t,t')$ and $\B(t,t')$. 
When the system is subject to a time-dependent rotation $\R(t)$ the deformation gradient and Finger 
tensors transform as \cite{larson}
\begin{eqnarray}
\hat{\E}(t,t') &=& \R(t)\E(t,t')\R^T(t'),
\\
\hat{\B}(t,t') &=& \R(t)\B(t,t')\R^T(t),
\end{eqnarray}
where the hat denotes a tensor in the rotated frame. 
Material objectivity is verified if the rotated stress tensor is found to be given by 
\begin{eqnarray}
\hat{\sig}(t) &=& \R(t)\sig(t)\R^T(t).
\label{material_objectivity}
\end{eqnarray}
A rather tedious but straightforward calculation shows that insertion of the 
rotated deformation tensors into our constitutive equation indeed leads to the relation (\ref{material_objectivity}).

\section{Distorted structure factor}\label{distorted_sec}
In order to calculate the distorted structure factor $S_{\kb}(t)$ we use the general 
integration-through-transients formula with $Q$-projected dynamics (\ref{average_QQQQ}) to 
calculate the average of a normalized product of two density fluctuations
\begin{eqnarray}
&&\hspace*{-0.cm}S_{\bf k}(t)
\!=\! \frac{1}{N}\langle \rho^*_{\bf k}\rho^{}_{\bf k} \rangle 
\\
&&\hspace*{0.6cm}+ \frac{1}{N}\int_{-\infty}^{t}\!\!dt' 
\langle \kap(t')\!:\!\hat{\boldsymbol{\sigma}}\,
Qe_-^{\int_{t'}^t ds\,Q\Omega^{\dagger}(s)Q}
Q\rho^*_{\bf k}\rho^{}_{\bf k} \rangle.
\notag
\end{eqnarray}
In order to arrive at a closed expression for $S_{\bf k}$ we approximate the average 
in the integrand using the time-dependent projection operator
(\ref{pair_projector_time}) 
\begin{eqnarray}
&&\hspace*{-0.6cm}\langle \kap(t')\!:\!\hat{\boldsymbol{\sigma}}\,
Qe_-^{\int_{t'}^t ds\,Q\Omega^{\dagger}(s)Q}
Q\rho^*_{\bf k}\rho^{}_{\bf k} \rangle
\\
&&\hspace*{-0.4cm}\approx
\langle \kap(t')\!:\!\hat{\boldsymbol{\sigma}}\,
QP_2(t,t')
e_-^{\int_{t'}^t ds\,Q\Omega^{\dagger}(s)Q}
P_2(t,t')Q\,
\rho^*_{\bf k}\rho^{}_{\bf k} \rangle.
\notag
\end{eqnarray}
The vertex function appearing to the left of the propagator is given by (\ref{trace_vertex}), 
albeit with advected wavevectors replacing static ones, whereas the vertex appearing on the 
right consists of two terms. 
The first of these terms is identical to (\ref{dyadic_vertex}), again with advected wavevectors 
replacing their static counterparts, whereas the second term is an isotropic contribution which 
we choose to neglect. Our choice to ignore this term is primarily motivated by the fact that it seems 
to be quantitatively small in comparison with the other terms \cite{fc09}. 
(We note, however, that the relative smallness of the isotropic term has been judged on the basis of  
calculations performed within the grand canonical ensemble \cite{henrich} which are not fully consistent 
with the requirement of fixed $N$ imposed by the Smoluchowski equation.) 
If we nevertheless choose to ignore this additional isotropic contribution and perform a factorization 
of the dynamical four point correlation function, we arrive at the result
\begin{eqnarray}
&&\hspace*{0.cm}S_{\bf k}(t)
=
S_k 
\\
&&+\int_{-\infty}^{t}\!\!\!\!dt'\,\left(
\frac{\kb(t,t')\cdot\kap(t')\cdot\kb(t,t')}{k(t,t')}
\right)
S'_{k(t,t')}\Phi^2_{\kb(t,t')}(t,t'),
\notag
\end{eqnarray}
%where $I_k(t)$ is an isotropic term given by 
%\begin{eqnarray}
%&&\hspace*{-0.6cm}I_k(t)=-\frac{S_{0}}{16\pi^3}\left(\frac{\partial S_k}{\partial \rho}\right)
%\int_{-\infty}^{t}\!\!\!\!dt'\,
%\int \!d\qb\;\times
%\\
%&&\hspace*{0.3cm}\times
%\left(
%\frac{\qb(t,t')\cdot\kap(t')\cdot\qb(t,t')}{q(t,t')}
%\right)
%\frac{S'_{q(t,t')}}{S^2_q}\Phi^2_{\qb(t,t')}(t,t'),
%\notag
%\end{eqnarray}
Simple application of the chain-rule enables us to then further simplify this expression 
to arrive at the final form
\begin{eqnarray}
S_{\kb}(t;\kap) = S_{k}\, -
\int_{-\infty}^t \!\!\!\!\!dt'\, \frac{\partial
S_{k(t,t')}}{\partial t'}\,\Phi^2_{\kb(t,t')}(t,t')
\label{distorted_structure}
%\\
%\!\!\!\!\!+&&\!\!\!\!\!\!\!
%\int_{-\infty}^t \!\!\!\!\!\!dt'\,
%\frac{\partial S_{k}}{\partial n}
%\!\!\int \!\!\frac{\!d\qb}{16\pi^3}
%\frac{\partial S_{q(t,t')}}{\partial t'}
%\frac{S_0}{S_q^2}\left( \!S_q \!+ n\frac{\partial S_q}{\partial n}\! \right)
%\Phi^2_{\!\qb(t,t')}(t,t'),
%\notag
\end{eqnarray}
This result has the appealing interpretation that in order to calculate the nonequilibrium 
flow-distorted structure factor one has simply to integrate the affinely advected 
equilibrium static structure factor over the flow history, weighted by the transient correlator encoding 
the structural relaxation of the system. 
We should also note that (\ref{distorted_structure}) makes clear the difference in philosophy between our 
approach and that of Miyazaki {\em et al.} \cite{miyazaki2004}. 
In \cite{miyazaki2004} the distorted stucture factor is employed as an input to the theory. 
In contrast, we input the equilibrium static structure factor, which serves as proxy for the potential 
interactions, and generate the structural distortion as an output. 

A notable aspect of our result (\ref{distorted_structure}) is that we can make a direct connection 
between the distorted structure factor and our constitutive equation (\ref{nonlinear}). 
By inspection, we find that substitution of (\ref{distorted_structure}) into the following 
expression 
\begin{eqnarray}
\sig(t)=-\frac{\rho}{16\pi^3}\int d\kb \,\frac{\kb\kb}{k} c'_k S_{\kb}(t)
\label{anisotropy_integral}
\end{eqnarray}
recovers (\ref{nonlinear}) exactly. 
For the case of shear flow (\ref{anisotropy_integral}) coincides with a result of Frerickson and 
Larson for copolymers \cite{fredrickson}, in some sense reflecting the Gaussian statistics 
underlying both approaches. 

From (\ref{anisotropy_integral}) it can be seen that our choice to ignore the isotropic 
term appearing in (\ref{distorted_structure}) corresponds to neglecting an isotropic contribution 
to the stress tensor. While the uncertainty regarding this term thus prevents us from predicting 
with confidence the system pressure, it has no consequence with regards the off diagonal stress tensor 
elements (shear stresses) or the normal stress differences which are, after all, the quantities 
of most direct rheological significance. 
Moreover, for incompressible systems the yielding behaviour of arrested states, as described by the 
yield stress surface \cite{pnas}, is invariant with respect to isotropic pressure. 
The nonequilibrium pressure becomes relevant when addressing flow-induced particle migration and 
shear banding \cite{besseling2010}. 
Finally, we note that the known ensemble dependence of the linear-response Green-Kubo 
result for the bulk viscosity of compressible fluids \cite{zwanzig_bulk} provides circumstantial 
evidence supporting our choice to suppress the isotropic contribution to (\ref{distorted_structure}).  
It seems likely that the choice of ensemble will be important when employing 
Green-Kubo-type formulae to calculate diagonal elements of the stress tensor.

\section{Discussion}\label{discussion}
We have presented a detailed derivation of our recently proposed constitutive equation for the 
rheology of dense colloidal suspensions \cite{brader2}. 
Our theory captures the slow structural relaxation arising from potential interactions (the `cage effect') 
by encoding these in the decay of the transient density correlator $\Phi_{\kb}(t)$. 
External flow leads to the affine advection of density fluctuations which competes with particle caging 
and tends to accelerate the loss of memory. It is this competition of timescales which is ultimately 
responsible for the observed transition between a shear thinning viscoelastic fluid and a yielding glass 
as a function of the coupling strength. 
The integration-through-transients formalism presented here provides a means to calculate stationary 
averages, correlation and response functions \cite{fc09,krueger} under arbitrary time-dependent flow. 

Our consitutive equation (\ref{nonlinear}) has been simplified in \cite{pnas} to a schematic level and 
solution of the resulting equation for a number of special cases has proved rather successful in 
capturing qualitative aspects of the stress response in a robust fashion. 
So far schematic calculations have been performed for steady flows \cite{pnas}, oscillatory shear 
\cite{brader_osc}, step strains \cite{pnas,thomas_step} and superposed shear and extensional flows \cite{farage}. 
Of particular note is that the schematic constitutive equation predicts a dynamic yield stress 
surface of von Mises form \cite{hill} together with small corrections related to the first normal stress 
difference. If the schematic model is indeed a faithful representative of the phenomenology presented 
by the full microscopic expression (\ref{nonlinear}), as currently seems to be the case, then these findings 
suggest that (\ref{nonlinear}) provides a possible route to calculating the yield stress surface from 
first-principles theory. In particular, the details of the yield surface could then be related directly 
to the potential interactions between the particles. 

A clear drawback of the present approach is the neglect of hydrodynamic interactions between the 
colloids. To a certain extent this may be defensible within the range of flow rates for which the basic 
assumptions of the theory remain valid, namely Peclet numbers $Pe\equiv \dot\gamma d^2/D_0$ less than 
around unity. 
Within this regime a simple modification of the bare diffusion coefficient may restore much of the 
quantitative error displayed by our Brownian dynamics-based  expressions when attempting to 
compare with experiment. 
However, such superficial incorporation of hydrodynamics will not affect the yield stress values 
predicted by the theory (occurring in the limit $\dot\gamma\rightarrow 0$) and may thus provide an 
incomplete picture. 
A more comprehensive method by which hydrodynamics can be included into mode-coupling-type theories 
is thus desirable (perhaps along the lines of \cite{banchio1,banchio2}), but care should be taken that 
modifications of the theory do not destroy the successful description of the hard-sphere glass 
transition.

Possibly the most questionable aspect of the present closure approximation is our reliance on the 
approximation $\Sigma=\overline{\Sigma}=0$ (\ref{big_sigma1},\ref{big_sigma2}), which constitutes a 
neglect of certain stress 
induced couplings, to arrive at a closed equation of motion for the transient density correlator. 
Other than the fact that they increase linearly with the accumulated strain, we know rather little 
about these flow induced terms and one can not draw on experience from the quiescent theory to assess 
their importance. 
The central nature of such an uncontrolled approximation within the present formulation could serve 
as motivation to seek an alternative formalism in which they do not occur (perhaps employing an 
operator other than ($\ref{omega_a}$).) 
In any case, if the yield strain were found to be large, then the implicit
assumption of a harmonic free energy functional, which underpins the entire
mode-coupling approach, would need correction by higher order terms; it
seems reasonable to hope that $\tilde{\Delta}_{\kb}(t,t')$ can be neglected on the same
basis. Some further arguments for neglecting $\tilde{\Delta}_{\kb}(t,t')$, made in the
context of steady shear flows but not specific to that case, are made in
\cite{fc09}. However further work to understand the origin of this term and if
possible to better justify its elimination would certainly be desirable. 

\section*{Acknowledgements}
We thank Th. Voigtmann and R.G. Larson for stimulating discussions. 
Work funded in part by EPSRC EP/E030173 and by the Deutsche Forschungsgemeinschaft in 
Transregio TR6. 
JMB acknowledges the support of the Swiss National Science Foundation. 
MEC holds a Royal Society Research Professorship.

\appendix
\section{Time ordered exponentials}\label{time_ordered}
The positively time-ordered exponential function of an arbitrary time-dependent operator 
$A(t)$ is defined by the series expansion \cite{vankampen,riskin} 
\begin{eqnarray}\label{eplus}
e_+^{\int_{t_1}^{t_2}ds A(s)}\!\!\!&=&\!
1 + \int_{t_1}^{t_2}\!\!\!ds_1 A(s_1) 
+ \int_{t_1}^{t_2}\!\!\!ds_1\!\int_{t_1}^{s_1}\!\!\!ds_2\, A(s_1)A(s_2)\notag\\
&&\hspace*{-1.5cm}+ \int_{t_1}^{t_2}\!\!\!ds_1\!\int_{t_1}^{s_1}\!\!\!ds_2
\!\int_{t_1}^{s_3}\!\!\!ds_3\, A(s_1)A(s_2)A(s_3)+\cdots.
\end{eqnarray}
The negatively ordered exponential is similarly defined by
\begin{eqnarray}\label{eminus}
e_-^{\int_{t_1}^{t_2}ds A(s)}\!\!\!&=&\!
1 + \int_{t_1}^{t_2}\!\!\!ds_1 A(s_1) 
+ \int_{t_1}^{t_2}\!\!\!ds_1\!\int_{t_1}^{s_1}\!\!\!ds_2\, A(s_2)A(s_1)\notag\\
&&\hspace*{-1.5cm}+ \int_{t_1}^{t_2}\!\!\!ds_1\!\int_{t_1}^{s_1}\!\!\!ds_2
\!\int_{t_1}^{s_3}\!\!\!ds_3\, A(s_3)A(s_2)A(s_1)+\cdots.
\end{eqnarray}
By multiplying out the series expansions in (\ref{eplus}) and (\ref{eminus}) the 
following identities can be proven 
\begin{eqnarray}\label{unity}
e_+^{\pm\int_{t_1}^{t_2}ds A(s)}e_-^{\mp\int_{t_1}^{t_2}ds A(s)}&=&1\\
e_-^{\pm\int_{t_1}^{t_2}ds A(s)}e_+^{\mp\int_{t_1}^{t_2}ds A(s)}&=&1,
\end{eqnarray}
where care must be taken that causality is respected within the multiple integrals.
These results are consistent with the expressions (\ref{def}) and (\ref{invdef}) for the 
deformation gradient and inverse, respectively. 
Using the series expansions (\ref{eplus}) and (\ref{eminus}) the following useful results for 
the derivatives of time-ordered exponential functions are easily proven
\begin{eqnarray}
\frac{\partial}{\partial t_2}\left(e_+^{\int_{t_1}^{t_2}ds A(s)}\right)
&=&\;\;A(t_2)\left(e_+^{\int_{t_1}^{t_2}ds A(s)}\right) \label{deriv1}\\
\frac{\partial}{\partial t_1}\left(e_+^{\int_{t_1}^{t_2}ds A(s)}\right)
&=&-\left(e_+^{\int_{t_1}^{t_2}ds A(s)}\right)A(t_1)\\
\frac{\partial}{\partial t_2}\left(e_-^{\int_{t_1}^{t_2}ds A(s)}\right)
&=&\;\;\;\left(e_-^{\int_{t_1}^{t_2}ds A(s)}\right)A(t_2)\label{deriv2}\\
\frac{\partial}{\partial t_1}\left(e_-^{\int_{t_1}^{t_2}ds A(s)}\right)
&=&-A(t_1)\left(e_-^{\int_{t_1}^{t_2}ds A(s)}\right)
\label{deriv4}
\end{eqnarray}
Finally we note that the adjoint operation satisfies 
\begin{eqnarray}
\left(e_+^{\int_{t_1}^{t_2}ds A(s)}\right)^{\dagger}
=e_-^{\int_{t_1}^{t_2}ds A^{\dagger}(s)}
\label{adjoint1}
\\
\left(e_-^{\int_{t_1}^{t_2}ds A(s)}\right)^{\dagger}
=e_+^{\int_{t_1}^{t_2}ds A^{\dagger}(s)}
\label{adjoint2}
\end{eqnarray}

\section{Operator identity}
The following operator identity has proved useful in analyzing the translational invariance 
of the two-time correlation functions
%\begin{eqnarray}\label{identity1}
%&&\hspace*{-0.7cm}e_+^{\,\int_{t_1}^{t_2}\!\!ds (A(s)+B(s))}
%=\notag\\
%&&\;\;\exp_+\left(\int_{t_1}^{t_2}\!\!ds\; e_+^{\int_s^{t_2}ds'A(s')}B(s)\,e_-^{-\int_s^{t_2}ds'A(s')}\right)
%\notag\\
%&&\;\;\times \exp_+\left(\int_{t_1}^{t_2}\!\!ds\; A(s)\right).
%\end{eqnarray}
\begin{eqnarray}\label{identity2}
&&\hspace*{-0.7cm}e_-^{\,\int_{t_1}^{t_2}\!\!ds (A(s)+B(s))}
=\exp_-\left(\int_{t_1}^{t_2}\!\!ds\; A(s)\right)\\
&&\;\;\times \exp_-\left(\int_{t_1}^{t_2}\!\!ds\; e_+^{-\int_s^{t_2}ds'A(s')}B(s)\,e_-^{\int_s^{t_2}ds'A(s')}\right)
\notag
\end{eqnarray}
where $A(t)$ and $B(t)$ are two arbitrary time-dependent operators.
The proof proceeds by first defining the operator
\begin{eqnarray}\label{U}
U^{+}(t_2,t_1)&=&e_-^{\int_{t_1}^{t_2}ds (A(s)+B(s))} \\
&=&\sum_{n=0}^{\infty}\int_{t_1}^{t_2}\!\!ds_1
\cdots\int_{t_1}^{s_{n-1}}\!\!\!\!ds_n(A(s_n)+B(s_n))\notag\\
&&\hspace*{3.0cm}\times\cdots(A(s_1)+B(s_1)),
\notag
\end{eqnarray}
where it is understood that the $n=0$ term is unity.
The derivative is thus given by
\begin{eqnarray}
\frac{\partial}{\partial t_2}U^{+}(t_2,t_1)=U^{+}(t_2,t_1)[A(t_2)+B(t_2)].
\label{comp1}
\end{eqnarray}
We now define a new operator 
\begin{eqnarray}\label{Utilde}
\tilde{U}(t_2,t_1)&=&e_-^{\int_{t_1}^{t_2}ds \,c(t_2,s)}\\
&=&\sum_{n=0}^{\infty}\int_{t_1}^{t_2}\!\!ds_1
\cdots\int_{t_1}^{s_{n-1}}\!\!\!\!ds_n c(t_2,s_n)\cdots c(t_2,s_1),\notag
\end{eqnarray}
where $c(t_2,s)$ is at present an arbitrary function. 
The derivative of this new operator is given by
\begin{eqnarray}
\label{Utilde_deriv}
\frac{\partial}{\partial t_2}\tilde{U}(t_2,t_1)&=&
\tilde{U}(t_2,t_1)c(t_2,t_2)\\
&&\hspace*{-2.4cm}+ \sum_{n=0}^{\infty}\int_{t_1}^{t_2}\!\!ds_1
\cdots\int_{t_1}^{s_{n-1}}\!\!\!\!ds_n\big[ c'(t_2,s_n)\cdots c(t_2,s_1)
\notag\\
&& \hspace*{-1cm}+ \;c(t_2,s_n)\,c'(t_2,s_{n-1})\cdots c(t_2,s_1)
\notag\\
&& \hspace*{0cm} \cdots 
\notag\\
&& \hspace*{-1cm}+ \;c(t_2,s_n)\cdots c'(t_2,s_2)\,c(t_2,s_1)
\notag\\
&& \hspace*{-1cm}+ \;c(t_2,s_n)\cdots c'(t_2,s_1) \big]
\notag
\end{eqnarray}
Defining $c(t_2,s)$ (up to boundary conditions) in the following way
\begin{eqnarray}
\frac{\partial}{\partial t_2}c(t_2,s) = 
c(t_2,s)A(t_2) - A(t_2)c(t_2,s) 
\label{c_equation}
\end{eqnarray}
leads to many convenient cancellations when substituted into (\ref{Utilde_deriv}), 
resulting in 
\begin{eqnarray}
\frac{\partial}{\partial t_2}\tilde{U}(t_2,t_1) &=& 
\tilde{U}(t_2,t_1)c(t_2,t_2) \\
&+& \tilde{U}(t_2,t_1)A(t_2) 
- A(t_2)\tilde{U}(t_2,t_1).\notag
\end{eqnarray}
Choosing the boundary condition $c(t_2,t_2)=B(t_2)$ yields
\begin{eqnarray}
\frac{\partial}{\partial t_2}\tilde{U}(t_2,t_1) + A(t_2)\tilde{U}(t_2,t_1) = 
\tilde{U}(t_2,t_1)(A(t_2)+B(t_2)). \notag\\
\label{temp}
\end{eqnarray}
Multiplying (\ref{temp}) on the right with a negatively ordered exponential yields 
\begin{eqnarray}
\frac{\partial}{\partial t_2}\left(e_-^{\int_{t_1}^{t_2}ds A(s)}\tilde{U}(t_2,t_1) \right)&=& \\
&&\hspace*{-4.0cm}\left(e_-^{\int_{t_1}^{t_2}ds A(s)}\tilde{U}(t_2,t_1) \right)(A(t_2)+B(t_2)). \notag
\label{comp2}
\end{eqnarray}  
Comparison of (\ref{comp1}) with (\ref{comp2}) allows the identification
\begin{eqnarray}
U^{+}(t_2,t_1)= e_-^{\int_{t_1}^{t_2}ds A(s)}\tilde{U}(t_2,t_1). 
\end{eqnarray}
Using (\ref{U}) and (\ref{Utilde}) to write $U^{+}$ and $\tilde{U}$ as exponentials 
thus yields
\begin{eqnarray}\label{almost}
e_-^{\int_{t_1}^{t_2}ds\,(A(s)+B(s))}= e_-^{\int_{t_1}^{t_2}ds\, A(s)}e_-^{\int_{t_1}^{t_2}ds \,c(t_2,s)}.
\end{eqnarray}
Finally we require an explicit form for $c(t_2,s)$. 
Using our chosen boundary condition $c(t_2,t_2)=B(t_2)$ enables solution of (\ref{c_equation}) 
\begin{eqnarray}\label{c_solution}
c(t_2,s)=e_+^{-\int_{t_1}^{t_2}ds'A(s')}B(s)e_-^{\int_{t_1}^{t_2}ds'A(s')},
\end{eqnarray}
as can be verified by substitution of (\ref{c_solution}) into (\ref{c_equation}). 
Substitution of (\ref{c_solution}) into (\ref{almost}) thus yields the desired result 
(\ref{identity2}). 

\section{Hadamard lemma}\label{hadamard}
For non-commuting, time-independent operators $X$ and $Y$ there exists a 
well-known identity (the Hadamard lemma)
\begin{eqnarray}\label{lemma}
e^{X}Ye^{-X}&=&Y \,+\, [X,Y] \,+\, \frac{1}{2!}[X,[X,Y]] \notag\\
&&\hspace*{-0.5cm}\,+\, \frac{1}{3!}[X,[X,[X,Y]]] \,+\, \cdots,
\end{eqnarray}
where $[X,Y]$ is the commutator.
Proof follows from defining the function $f(s)=e^{sX}Ye^{-sX}$, Taylor expanding 
about $s=0$ and then setting $s=1$.
(\ref{lemma}) can be particularly useful in cases for which the sequence of nested 
commutators truncates at a low order, or when the infinite series can be resummed into 
a closed form. 

When employing the operator identity (\ref{identity2}) a structure 
analogous to the left hand side of (\ref{lemma}) arises naturally in one of the factorized exponential 
functions. 
We are thus motivated to find a continuous version of the Hadamard lemma, which may be useful 
in simplifying (\ref{identity2}) for certain special cases. 
For arbitrary time-dependent operators $A(t)$ and $B(t)$, we find the following nested commutator 
expansion
\begin{eqnarray}\label{hadamard1} 
e_+^{\!-\int_s^{t}ds'A(s')}B(s)\,e_-^{\int_s^{t}ds'A(s')} \hspace{-0.2cm}&=& 
\!B(s) -\! \int_s^t \!\!ds_1\, [A(s_1),B(s)] \notag\\
&&\hspace*{-4cm}+ \int_s^t \!\!ds_1 \int_s^{s_1} \!\!\!ds_2 \,[A(s_1),[A(s_2),B(s)]] \notag\\
&&\hspace*{-4cm}- \int_s^t \!\!ds_1 \int_s^{s_1} \!\!\!ds_2 \int_s^{s_2} \!\!\!ds_3  \,[A(s_1),[A(s_2),[A(s_3),B(s)]]] 
\vspace*{0cm}\notag\\
\notag\\
&&\hspace*{-4cm}+ \cdots,
\end{eqnarray}
%\begin{eqnarray}\label{hadamard2} 
%e_+^{\int_s^{t}ds'A(s')}B(s)\,e_-^{-\int_s^{t}ds'A(s')}\hspace{-0.2cm}&=& 
%\!B(s) +\! \int_s^t \!\!ds_1\, [A(s_1),B(s)] \notag\\
%&&\hspace*{-4cm}+ \int_s^t \!\!ds_1 \int_s^{s_1} \!\!\!ds_2 \,[A(s_1),[A(s_2),B(s)]] \notag\\
%&&\hspace*{-4cm}+ \int_s^t \!\!ds_1 \int_s^{s_1} \!\!\!ds_2 \int_s^{s_2} \!\!\!ds_3  \,[A(s_1),[A(s_2),[A(s_3),B(s)]]] 
%\vspace*{0cm}\notag\\
%\notag\\
%&&\hspace*{-4cm}+ \cdots
%\end{eqnarray}
which is a continuum version of (\ref{lemma}). 
The proof proceeds by first defining the operator
\begin{eqnarray} 
F(t,s)=e_+^{\!-\int_s^{t}ds'A(s')}B(s)\,e_-^{\int_s^{t}ds'A(s')}.
\end{eqnarray}
Differentiation with respect to $t$ leads to 
\begin{eqnarray}\label{partialderiv} 
\frac{\partial F(t,s)}{\partial t} = -A(t)F(t,s) + F(t,s)A(t),
\end{eqnarray}
where we have used (\ref{deriv1}) and (\ref{deriv2}). 
Integration of (\ref{partialderiv}) from $s$ to $t$ yields a recursion relation
\begin{eqnarray}\label{recursion}
F(t,s) =  B(s) - \!\!\int_s^t \!\!\!ds_1  (A(s_1)F(s_1,s) \!-\! F(s_1,s)A(s_1)), 
\notag\\
\end{eqnarray} 
where we have made the identification $F(s,s)=B(t)$. Straightforward iteration 
of (\ref{recursion}) directly generates the series (\ref{hadamard1}).

\section{Volterra equation}\label{volterra}
A Volterra integral equation of the second kind for the function $\Phi_{\kb}(t,t')$ can be 
written the following form \cite{tricomi}
\begin{eqnarray}\label{volt1}
\Phi_{\kb}(t,t') = g_{\kb}(t,t') + \int_{t'}^t \!ds\,K_{\kb}(t,s,t')\,\Phi_{\kb}(s,t').
\end{eqnarray}
The wavector subscripts are irrelevant in the formal solution of this temporal integral 
equation, but are nevertheless included here to facilitate comparison with the equations 
appearing in the main text. 
Subject to certain reasonable conditions on both the kernel $K_{\kb}(t,s,t')$ and inhomogeneity 
$g_{\kb}(t,t')$, the solution is given by 
\begin{eqnarray}\label{volt2}
\Phi_{\kb}(t,t') = g_{\kb}(t,t') - \int_{t'}^t \!ds\,R_{\kb}(t,s,t')\,g_{\kb}(s,t'),
\end{eqnarray}
where the resolvent kernel $R(t,s,t')$ satisfies the following linear 
integral equation 
\begin{eqnarray}\label{v3}
R_{\kb}(t,s,t') &=& -K_{\kb}(t,s,t') 
\\
&+& \int_{s}^t \!\!ds'\,K_{\kb}(s',s,t')\,R_{\kb}(t,s',t').
\notag
\end{eqnarray}
These results can be used to solve directly the inhomogeneous equation (\ref{inhom}). 
Comparing (\ref{inhom}) with (\ref{volt1}) enables the identification 
\begin{eqnarray}\label{v4}
K_{\kb}(t,s,t')=-M_{\bf k}(t,s,t')/\Gamma_{\bf k}(t,t'). 
\end{eqnarray}
Substitution of (\ref{v4}) into (\ref{v3}) and comparison of the resulting equation with (\ref{Mm_equation}) 
identifies the resolvent kernel 
\begin{eqnarray}\label{v5}
R_{\kb}(t,s,t')=-m_{\kb}(t,s,t')\Gamma_{\kb}(s,t').
\end{eqnarray}
Substitution of (\ref{v5}) into (\ref{volt2}) yields the solution (\ref{second_eom}).

\section{Simplifying the friction kernel}\label{simplifying}
When attempting to simplify the friction kernel entering (\ref{second_eom}) it is useful to 
convert advection operators involving $\delta\Omega^{\dagger}(t)$  
to analogous quantities involving $\overline{\delta\Omega^{\dagger}}(t)$. 
This can be achieved using the identities
\begin{eqnarray}\label{replacement1}
e_-^{\int_{t'}^t \!ds\, \delta\Omega^{\dagger}(s)} 
&=& 
e_-^{\int_{t'}^t \!ds\, \overline{\delta\Omega^{\dagger}}(s)} 
\Big(
1 + \Sigma(t,t')
\Big)
\\
\label{replacement2}
e_+^{-\int_{t'}^t \!ds\, \delta\Omega^{\dagger}(s)} 
&=& 
\Big(
1 + \overline{\Sigma}(t,t')
\Big)
e_+^{-\int_{t'}^t \!ds\, \overline{\delta\Omega^{\dagger}}(s)} 
\end{eqnarray}
where we have introduced the two operators
\begin{eqnarray}\label{big_sigma11}
&&\hspace*{-0.9cm}
\Sigma(t,t') \!=\!\!
\int_{t'}^t\!\!ds \, e_+^{-\int_{s}^t \!ds'\, \overline{\delta\Omega^{\dagger}}(s')}
\kap^{T}(s)\!:\!\hat{\sig}\,e_-^{\int_{s}^t \!ds'\, \delta\Omega^{\dagger}(s')} 
\\
\label{big_sigma22}
&&\hspace*{-0.9cm}
\overline{\Sigma}(t,t') \!= \!\!
\int_{t'}^t\!\!ds \, e_+^{-\int_{s}^t \!ds'\, \delta\Omega^{\dagger}(s')}
\kap^{T}(s)\!:\!\hat{\sig}\,e_-^{\int_{s}^t \!ds'\, \overline{\delta\Omega^{\dagger}}(s')}.
\end{eqnarray}
In the main text we will argue that both $\Sigma$ and $\overline{\Sigma}$ can 
be set to zero in the final approximation. 
Proof of (\ref{replacement1}) follows from considering the derivative 
\begin{eqnarray}
&&\frac{\partial}{\partial t'}\left(
e_+^{-\int_{t'}^t \!ds\, \overline{\delta\Omega^{\dagger}}(s)}
e_-^{\int_{t'}^t \!ds\, \delta\Omega^{\dagger}(s)} 
\right)
=
\\
&&\hspace*{1cm}e_+^{-\int_{t'}^t \!ds\, \overline{\delta\Omega^{\dagger}}(s)}
\left(
\overline{\delta\Omega^{\dagger}}(t) - \delta\Omega^{\dagger}(t)
\right)
e_-^{\int_{t'}^t \!ds\, \delta\Omega^{\dagger}(s)} 
\notag
\end{eqnarray}
and integration from $t'$ to $t$. 
The proof of (\ref{replacement2}) is analogous.

The identity (\ref{replacement1}) is helpful in simplifying the irreducible time evolution 
operator given by 
\begin{eqnarray}
\mathcal{G}^{\rm irr}(t,t')=\mathcal{G}(t,t')\tilde{Q}(t,t'). 
\end{eqnarray}
Using (\ref{replacement1}) to replace the leftmost ordered exponential in (\ref{g_operator}) 
yields a natural division into two terms
\begin{eqnarray}
\mathcal{G}^{\rm irr}(t,t')=\big(\mathcal{G}_Q(t,t') + \mathcal{G}_{\Sigma}(t,t')\big)\tilde{Q}(t,t'), 
\end{eqnarray}
where
\begin{eqnarray}
\label{GQ}
&&\hspace*{-0.9cm}\mathcal{G}_Q(t,t')\!=e_{-}^{\int_{t'}^t \!ds\,\overline{\delta\Omega^{\dagger}}(s)}
Q(t,t')\,\Omega_e^{\dagger}\,
e_{+}^{-\!\int_{t'}^t \!ds\,\delta\Omega^{\dagger}(s)}
\\
&&\hspace*{-0.9cm}\mathcal{G}_{\Sigma}(t,t')\!=e_{-}^{\int_{t'}^t \!ds\,\overline{\delta\Omega^{\dagger}}(s)}
\Sigma(t,t')Q(t,t')\,\Omega_e^{\dagger}\,
e_{+}^{-\!\int_{t'}^t \!ds\,\delta\Omega^{\dagger}(s)}
\end{eqnarray}
The purpose of this manipulation is that under the assumption that $\Sigma(t,t')$ vanishes 
we obtain $\mathcal{G}^{\rm irr}(t,t')\approx \mathcal{G}_Q(t,t')\tilde{Q}(t,t')$, which is an operator in the 
space perpendicular to linear density fluctuations. 
This consideration then leads us to define the propagator 
\begin{eqnarray}\label{perpendicular_dynamics}
U^{\rm irr}_Q(t,s,t')\equiv e_-^{\int_{s}^{t}\!ds'\,\mathcal{G}_Q(s',t')\tilde{Q}(s',t')}
\end{eqnarray}
which generates dynamics in the space perpendicular to linear density fluctuations.

\end{document}